\documentclass[10pt,twocolumn,preprintnumbers,aps,amsmath,amssymb,prd,nofootinbib
,superscriptaddress]{revtex4-1}
\usepackage{graphicx,longtable,mathrsfs,color,array}
\usepackage[hidelinks]{hyperref}
\usepackage[usenames,dvipsnames]{xcolor} 
\usepackage{amssymb,amsmath,mathtools,mathrsfs,slashed} 
\usepackage{epsfig,subfigure,float} 
\usepackage{booktabs,longtable,ctable,multirow} 
\usepackage{exscale,relsize} 
\usepackage[normalem]{ulem} 
\usepackage{enumerate}
\usepackage{times, mathptmx} 
\usepackage[utf8]{inputenc}
\usepackage{color}
\usepackage{hyperref}
\usepackage{graphicx}
\usepackage{color}
\usepackage{physics}
\usepackage{graphicx,graphics}
\usepackage{bm}
\usepackage{tabularx}
\allowdisplaybreaks[1]
\usepackage[section]{placeins}

\newcommand{\eV}{\textup{eV}}
\newcommand{\pc}{\textup{pc}}

\begin{document}

\title{Growth of accretion driven scalar hair around Kerr black holes}
\author{Jamie Bamber}
\email{james.bamber@physics.ox.ac.uk}
\affiliation{Astrophysics, University of Oxford, DWB, Keble Road, Oxford OX1 3RH, UK}
\author{Katy Clough}
\email{katy.clough@physics.ox.ac.uk}
\affiliation{Astrophysics, University of Oxford, DWB, Keble Road, Oxford OX1 3RH, UK}
\author{Pedro G. Ferreira}
\email{pedro.ferreira@physics.ox.ac.uk}
\affiliation{Astrophysics, University of Oxford, DWB, Keble Road, Oxford OX1 3RH, UK}
\author{Lam Hui}
\email{lhui@astro.columbia.edu}
\affiliation{Department of Physics and Astronomy, Columbia University, New York, NY 10027, USA}
\author{Macarena Lagos}
\email{mal2346@columbia.edu}
\affiliation{Department of Physics and Astronomy, Columbia University, New York, NY 10027, USA}

\date{Received \today; published -- 00, 0000}

\begin{abstract}
Scalar fields around compact objects are of interest for scalar-tensor theories of gravity and dark matter models consisting of a massive scalar, e.g.~axions. We study the behaviour of a scalar field around a Kerr black hole with non trivial asymptotic boundary conditions -- both non zero density and non zero angular momentum. Starting from an initial radially homogeneous configuration, a scalar cloud is accreted, which asymptotes to known stationary configurations over time.
We study the cloud growth for different parameters including black hole spin, scalar field mass, and the scalar field density and angular momentum far from the black hole. We characterise the transient growth of the mass and angular momentum in the cloud, and the spatial profile of the scalar around the black hole, and relate the results of fully non-linear simulations to an analytic perturbative expansion. We also highlight the potential for these accreted clouds to create monochromatic gravitational wave signals - similar to the signals from superradiant clouds, although significantly weaker in amplitude.

\end{abstract}

\maketitle


\section{Introduction}

Black holes are some of the most exotic objects in the Universe, describing strong gravity environments and testing the limits of General Relativity (GR). Recent advances in observational astronomy, including the observation of the horizon of the black hole (BH) at the centre of $M87$ by the Event Horizon Telescope \cite{EHT}, and the launch of gravitational wave astronomy with Advanced LIGO \cite{LIGO}, VIRGO \cite{VIRGO}, KAGRA \cite{KAGRA} and in future LISA \cite{LISA} and other detectors \cite{ET, CE}, allow us to probe these objects in ever finer detail. 

Despite their exotic nature, a set of uniqueness theorems  -- dubbed ``no-hair" theorems -- state that the properties of black holes can be fully determined by a small number of parameters: their mass $M$, angular momentum $J$, and electric charge $Q$, and cannot support non-trivial profiles of other fields \cite{Carter_1971, Israel_1968, NoHair_Heusler_1996, Asym_BH_review}. These theorems apply under both General Relativity and a broad class of extensions which are of cosmological interest \cite{NoHair_Galileon, NoHair_noncan, Sotiriou_2015}. However these theorems depend on a number of restrictive assumptions, and violating one or more of these may allow black holes to acquire hair \cite{Jacobson_1999,NoHair_noncan,Asym_BH_review}. New fundamental fields are common feature of theories of beyond the Standard Model physics \cite{BSM_review} and theories of modified gravity \cite{Clifton_2012, Horndeski, Galileon}. Investigating situations in which hair could grow in the extreme environment of a black hole is therefore strongly motivated as a way of detecting new physics.

\begin{figure}[ht]
    \centering
    \includegraphics[width=0.45\textwidth]{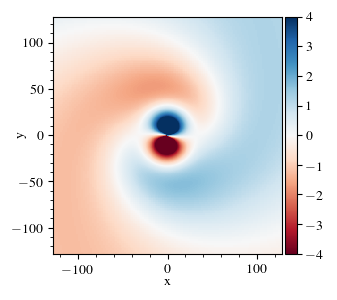}
    \caption{Visualisation of the 2D scalar field profile $\varphi/\Phi_0$ around a spinning BH, after a time $t=800M$. The initial scalar profile at $t=0$ is set as a $l=m=1$ spherical harmonic. The scalar and BH masses are given by $\mu=0.4$ and $M=1$ respectively, and the dimensionless BH spin is set to $\chi=0.7$ (aligned with the scalar spin axis). At this point the maximum amplification of the scalar field is of order $|\varphi/\Phi_0|\sim 10$, where $\Phi_0$ is the initial amplitude, but this will continue to grow over time.
    }
    \label{fig:spiral}
\end{figure}

Arguably, the simplest case of an extra field one might consider is a real scalar field $\varphi$ with mass $\mu$ and a potential $V(\varphi) \approx \tfrac{1}{2}\mu^2\varphi^2$ minimally coupled to Einstein gravity
\footnote{Throughout this paper we use geometric units $G=c=1$, in which the scalar field is parameterised by the inverse reduced Compton wavelength $\mu=\lambdabar_c^{-1}$ or the dimensionless parameter $\alpha_g = \mu M$. We sometimes refer to $\mu$ as the mass, although the physical mass is given by $m_s = \hbar\mu$. (In units where M is the BH mass, $\hbar\neq1$).}
. This potential could also be taken as the leading order expansion around a minimum of a more general potential. The discovery of the Higgs boson confirms the existence of such fundamental scalar fields  in nature \cite{Higgs_ATLAS,Higgs_CMS} and well motivated proposed fields include the QCD axion \cite{BSM_review}, with masses in the range $10^{-5}\eV$ to $10^{-3}\eV$ \cite{Review_Particle_Physics_2019}, and axions resulting from string theory which can have a wide range of masses \cite{Axions_string_theory, String_axiverse}.
Dark matter (DM) composed of ultra-light bosons with masses of order $10^{-22}\eV$, so called ``fuzzy dark matter", has been proposed as a potential solution for discrepancies between simulation and observation, as it results in modified structure formation on small cosmological scales \cite{Hu_2000, Sikivie_2011, Schive:2014dra, Hui_2017, Bar_2018, Hui:2020hbq}. Due to its high occupation number, is well described by a massive classical scalar field \cite{ferreira2020ultralight} (see \cite{Marsh:2015xka,Review_Scalar_Field_DM,Niemeyer:2019aqm} for reviews).

Analytic solutions for the stationary profile of scalar clouds have been obtained for a massive scalar field obeying the Klein-Gordon equation on fixed BH background metric \cite{Kerr_sols, Hui_2019} (where the backreaction of the scalar field on the metric is neglected) based on the confluent Heun function \cite{Heun}. For $r \gg M$ the scalar field obeys an equation which approximates the Schr{\"o}dinger equation governing the electron in a hydrogen atom \cite{Grav_atom}. Hence in this non-relativistic limit the field profiles become energy levels analogous to those of hydrogen.

A key question is the likelihood of dynamical formation of such scalar clouds. One phenomenon that has generated much interest is their growth around a rotating BH via superradiance. This mechanism allows a Kerr black hole of mass $M$ to transfer energy to a rotating massive boson cloud, provided the frequency of the field $\omega$ satisfies the instability condition 
\begin{equation}
M\omega < m\Tilde{\Omega}_H,
\label{eq:Sup}
\end{equation}
where $m$ is the azimuthal mode number of the field and $\Tilde{\Omega}_H$ is the dimensionless angular velocity of the BH horizon, given by $a/(2 r_+)$ where $a$ is the ratio of black hole angular momentum to mass and $r_+$ is the horizon radius in Boyer-Linquist coordinates \cite{Superradiance_Review, Arvanitaki_2015, East_2017, Witek_2013}. Since $\omega \sim \mu$, bosonic fields with appropriate masses can grow, even from small quantum fluctuations, on short astrophysical timescales. On the other hand, superradiance requires a fairly tightly tuned resonance between particle and BH length scales, which may not be realised in nature.\footnote{The corresponding mass scales are $10^{-11}$eV for a typical astrophysical BH mass $10 M_{\odot}$, and $10^{-19}$eV for a supermassive BH of mass $10^9 M_{\odot}$ \cite{Hui_2019}.}

By moving beyond the superradiant mechanism and its requirements we can consider how scalar hair could form under a broader class of conditions. As shown by Jacobson \cite{Jacobson_1999}, giving the field a non-trivial time dependence far from the black hole is one way of violating the assumptions of the no-hair theorems and gives rise to the growth of a non-trivial profile. In \cite{Clough_2019, Hui_2019} the Jacobson effect was explored as an alternative, simpler mechanism for growing scalar hair.
In flat space the scalar field can have a global non trivial time dependence of the form $e^{-i\mu t}$ \cite{Hui_2019}. We can therefore impose this as the asymptotic condition for the field, rather than requiring it to decay to zero as in superradiant bound states.  The physical interpretation of this condition is that we have a non zero asymptotic density - that is, the clouds arise from simple gravitational accretion from a cosmological background. Unlike superradiance, it does not rely on a particular value of $M\mu$, but wave like effects will be most pronounced in the regime $M\mu \lesssim 1$, which is particularly relevant to the case of light bosonic dark matter.

\begin{figure}
    \centering
    \subfigure[$\;$ Scalar field energy density $\rho_E/\rho_0$ \label{fig:KBH_rho}]{
           \includegraphics[width=0.4\textwidth]{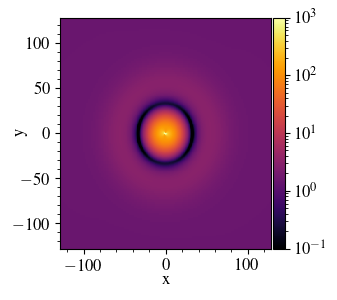}}
    \subfigure[$\;$ Scalar field energy density $\rho_E/\rho_0$ (zoomed in) \label{fig:rho_close_up}]{
           \includegraphics[width=0.4\textwidth]{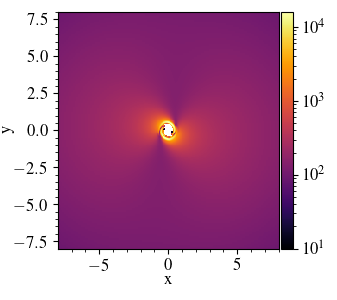}}
    \subfigure[$\;$Scalar field angular momentum density magnitude $\vert \rho_J \vert/\rho_0$  \label{fig:KBH_rho_azimuth}]{
           \includegraphics[width=0.4\textwidth]{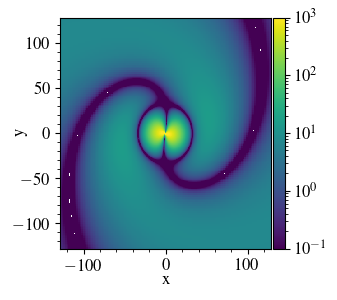}}
    \caption{Visualisation of the energy density $\rho_E/\rho_0$ and angular momentum density $\rho_J/\rho_0$ profiles for the same parameters and at the same time as in Fig. \ref{fig:spiral} of $l=m=1, \mu=0.4, M=1, \chi=0.7, t=800$, on a logarithmic scale. The initial density $\rho_0 = \tfrac{1}{2}\Phi^2_0\mu^2$ where $\Phi_0$ is the initial amplitude of the field. Note that for a real scalar field with non zero angular momentum there is a non axisymmetric, rotating component to both energy and angular momentum densities.}
    \label{fig:energy}
\end{figure}

The numerical simulations in \cite{Clough_2019} demonstrated the growth of a non-trivial profile with and without backreaction on the metric, assuming a spherically symmetric Schwarzschild BH and DM environment. We now extend this work to consider the impact of angular momentum on the accretion of the scalar field, including the interplay of both non zero asymptotic angular momentum in the scalar field and a spinning Kerr black hole. We study a range of scalar masses, characterising deviations from uniform spherically symmetric accretion, and their impact on the cloud growth. Fig.~\ref{fig:spiral} and Fig.~\ref{fig:energy} illustrate one example of the resulting clouds. Astrophysically, the non-zero asymptotic angular momentum may arise from rotating galactic DM halos or from the merger of a compact object binary where each object has its own scalar cloud.

This work is relevant for isolated astrophysical BHs, and therefore constraints on scalar hair from imaging by the EHT and the motion of bulge stars \cite{Cunha:2019ikd, Brax:2019npi, Khodadi:2020jij, Boskovic:2018rub, De_Martino_2020}. It may also shed light on the end state of the scalar field around the remnant of a binary merger. The interaction of minimally coupled light bosonic fields with compact binaries has been examined in analytic and numerical studies by a number of authors \cite{Horbatsch_2011, Berti_2013, Blas_2017, Ferreira_M_2017, Berti_2019, Baumann_2019, Wong_2019, Wong_2020, Rozner_2020, annulli2020, Annulli_2020_PRD, Ikeda:2020xvt}.
As pointed out in \cite{Hui_2019}, in the limit of high scalar masses, the configurations we study correspond to the dark matter spikes of \cite{Gondolo:1999ef} for which the dephasing of the gravitational wave signal has recently been studied in \cite{Kavanagh_2020}. Our own simulations will be extended to explicitly study the binary case in a separate work.

The paper is organised as follows. The formalism and setup for our numerical work is described in Sec. \ref{sec-methods}, with further detail in Appendix \ref{App:Flux}. In Sec.\ref{sec-analytic} we develop a perturbative analytic framework to describe the accretion onto the BH as a function of the various parameters. We then confirm these predictions and compare them to the full non linear evolution by performing simulations as described Sec. \ref{sec-numerics}. In Sec. \ref{sec-gws} we highlight the potential for our scalar clouds with angular momentum to generate continuous monochromatic gravitational wave signals, and quantify their amplitude. In Sec \ref{sec-discussion} we summarise our findings and propose directions for future work.

\section{Framework and Numerical Setup}
\label{sec-methods}
In this section we lay out the formalism and methods we will use to study the cloud growth.

\subsection{Kerr metric background}

In Boyer Lindquist (BL) coordinates \{$t$, $r$, $\theta$, $\phi$\} the spacetime line element of a Kerr black hole is given by \cite{Kerr_sols}:
\begin{equation}
\begin{split}
	\dd s^2 = - \left(   1 - \frac{2Mr}{\Sigma} \right) \dd t^2 - \frac{4 a M r \sin^2\theta}{\Sigma} \dd t \dd \phi \\
	+ \frac{\Sigma}{\Delta} \dd r^2 + \Sigma \dd \theta^2 + \frac{\mathcal{A}}{\Sigma} \sin^2\theta \dd \phi^2,
	\label{KerrBL}
\end{split}
\end{equation}
where 
\begin{align}
	\mathcal{A} &= (r^2 + a^2)^2 - \Delta a^2 \sin^2 \theta ~, \\
	\Delta &= r^{2}-2Mr+a^{2}, \\
	\Sigma &= r^{2}+a^{2}\cos ^{2}\theta,
\end{align}
and	$a=J/M$ is the Kerr spin parameter, with $J$ being the angular momentum of the BH, and $M$ the mass of the black hole. We can also define the dimensionless spin parameter  $\chi = a/M$ which takes a value between 0 (Schwarzschild) and 1 (extremal spin). Notice that the BH is entirely parameterised by $\chi$ and $M$, where the alignment of the angular momentum is taken to be in the $z$ direction, without loss of generality. The inner and outer event horizons of the black hole are located at $r_{\pm} = M \pm \sqrt{M^2 - a^2}$. In the simulations performed in this paper, we use the ``Quasi-Isotropic Kerr" (QIK) coordinates of \cite{Liu:2009al, Brandt:1994ee, Brandt:1996si, Okawa:2014nda}. These are obtained by introducing the quasi-isotropic radial coordinate $R$, related to the BL radius $r$ via
\begin{equation}
	r = R \left( 1 + \frac{r_+}{4R} \right)^2 ~. \label{eqn:Randr} 
\end{equation}
(Note that we use a Cartesian realisation of these coordinates where $R^2 = x^2 + y^2 + z^2$ and $\{x,y,z\}$ are related to $\{R,\theta,\phi\}$ in the usual Cartesian manner.) In QIK coordinates $\{t,R,\theta,\phi\}$, the line element is
\begin{equation}
\dd s^2=-\alpha^2\,\dd t^2+\gamma_{ij}(\dd x^i + \beta^i\,\dd t)(\dd x^j + \beta^j\,\dd t),
\end{equation}
where 
\begin{align}
	&\alpha = \sqrt{\frac{\Delta \Sigma}{\mathcal{A}}},\quad 	\beta^\phi = -\frac{2 a M r}{\mathcal{A}},\nonumber\\
	& 	\gamma_{ij}\dd x^i \dd x^j = \frac{\Sigma r}{R^2(r - r_-)} \dd R^2 + \Sigma \dd \theta^2 + \frac{\mathcal{A}}{\Sigma} \sin^2 \theta \dd \phi^2,
\end{align}
and the other $\beta^i$ components vanish. 
The advantage of this is that the coordinate location of the outer horizon is fixed at $R = R_+ = r_+/4$, which maintains a finite value $M/4$ in the extremal spin limit. One difficulty is that the lapse $\alpha$ goes to zero at the horizon, so they are not horizon penetrating. We use an analytic continuation such that inside the horizon the lapse becomes 
\begin{equation}
	\alpha = -\sqrt{\left\vert\frac{\Delta \Sigma}{\mathcal{A}}\right\vert}.
\end{equation}
Further details regarding this choice and a comparison to horizon penetrating Kerr-Schild coordinates are provided in Appendix \ref{App:Flux}. 

\subsection{Scalar field evolution and initial conditions}
\label{sec-methods-IC}
We will consider a minimally-coupled massive scalar field $\varphi$ with mass scale $\mu$. The equation of motion for the field is the Klein-Gordon equation
\begin{equation}
    \left[-\nabla^\nu \nabla_\nu + \mu^2\right]\varphi=0 ~.
\end{equation}
Throughout this work we will neglect backreaction of the scalar field on the metric, which is a very good approximation for low density fields. As shown in \cite{Clough_2019}, the inclusion of backreaction does not disrupt the accretion process. 
In particular, this is a reasonable approximation for typical DM densities. The backreaction will be order $G\rho$ where $\rho$ is the scalar field energy density. For a typical DM density of $\rho \sim 1~\textup{GeV}\textup{cm}^{-3}$ \cite{Pato_2015} the source term expressed in geometric units where $M=1$ is $G\rho \sim 10^{-30} (M/10^6M_{\odot})^2$ which is $\lll 1$ even for supermassive BHs. 

In the absence of the black hole's potential well a spatially homogeneous massive scalar simply oscillates at frequency $\mu$ as $\varphi(t) = \Phi_0 e^{-i\mu t}$. This corresponds to the case of zero angular momentum in the field, and spatially constant energy density. Adding in angular variation, in particular a spherical harmonic profile, corresponds to adding a non zero angular momentum to the field in some direction.
This observation motivates the choice of our initial conditions as
\begin{equation}
    \varphi(t,\theta,\phi) = \Phi_0 \textup{Re}\{e^{-i\mu t} Y^*_{lm}(\theta, \phi)\},
\end{equation}
where $Y_{lm}$ are spherical harmonics. We consider different values for the dimensionless spin of the BH $\chi$, the dimensionless ratio of the BH radius versus the scalar wavelength $M\mu$, the mode numbers $l,m$ of the initial angular momentum of the surrounding scalar field, and the alignment angle $\alpha$ between the black hole and cloud spin. 

The maximum size of the cloud that develops will be strongly influenced by the surrounding scalar environment. Following \cite{Hui_2019} one can define a ``radius of influence" of the black hole as the radius at which the virial velocity calculated from the black hole's potential is comparable to the typical velocity dispersion of the surrounding scalar matter,
\begin{equation}
    v^2_{\textup{virial}} = \frac{M}{r_i} \sim v^2_{\textup{disp}}.
\end{equation}
The conditions at this radius will determine the characteristics of the cloud that form around the BH. Note that this is different from the superradiant case, where the scalar field is assumed to decay to zero at large $r$ and the size of the cloud is determined by the properties of the black hole. In that case the characteristic size is given by \cite{Brito_2015}
\begin{equation}
    R \sim M/(M\mu)^2.
\end{equation}

In this work we remain agnostic to the exact conditions, and simply characterise the cloud according to the possible physical parameters of the asymptotic scalar distribution. However, here we will briefly illustrate the physical intepretation of the quantities in the dark matter case. The energy density of the field is $\rho_E \sim \Phi^2_0\mu^2$. Thus for a typical DM density of $\sim 1\textup{GeVcm}^{-3}$ and a supermassive BH of mass $10^6 M_{\odot}$ we would have $ \Phi_0 \sim 10^{-15} (\mu M)^{-1}$ in geometric units where $M=1$. In our simulations the value of $\Phi_0$ is arbitrary as we neglect backreaction, so we use an order 1 value which can be rescaled accordingly for different physical densities. Whilst our initial configuration is somewhat artificial, corresponding to a homogeneous radial density profile, it is nevertheless instructive in showing how different parameters of the field and BH affect the transient growth rate from a zero cloud state, and allows us to relate the change in the mass and spin of the BH to the non trivial asymptotic conditions which characterise the scalar field far from the BH.

\subsection{Diagnostic quantities}
\label{sec:diagnostics}
In this section we define a number of quantities that will allow us to quantify the growth of the scalar cloud and its effects. Further implementation details are given in Appendix \ref{App:Flux}.

The Kerr metric is independent of $t$ and $\phi$ and so admits two Killing vectors $\xi^{\mu}_1 = (1, 0, 0, 0)$ and $\xi^{\mu}_2 = (0, 0, 0, 1)$ in $(t, R, \theta, \phi)$ coordinates, with associated conserved quantities.
The properties of a Killing vector field and the energy-momentum tensor then imply
\begin{equation}
    \nabla_{\mu}(\xi^{\nu}T^{\mu}_{\nu}) = \frac{1}{\sqrt{-g}}\partial_{\mu}(\sqrt{-g}\xi^{\nu}T^{\mu}_{\nu}) = 0,
\end{equation}
where $g$ is the determinant of the metric $g_{\mu\nu}$, and therefore we can define two associated conserved currents as: 
\begin{align}
    J^{\mu}_t &= -T^{\mu}_t, \label{tcurr}\\
    J^{\mu}_{\phi} &= T^{\mu}_{\phi}.
\end{align}
which obey $\nabla_{\mu}J^{\mu} = 0$ and
\begin{align}
    \partial_t (\sqrt{-g}J^t) &= - \partial_i (\sqrt{-g} J^i).
    \label{eq:J_cons}
\end{align}
Note that $\sqrt{-g} = \alpha \sqrt{\gamma}$ where $\gamma$ is the determinant of the spatial metric $\gamma_{ij}$. Integrating both sides of \eqref{eq:J_cons} over a 3D spatial volume $\Sigma$ within one spatial slice and applying the divergence theorem gives
\begin{align}
    \partial_t ~ \int_{\Sigma} \sqrt{-g} J^t \dd x^3 &= - \int_{\partial \Sigma} \sqrt{-g} J^i \dd S_i, \\
    \partial_t ~ \int_{\Sigma} \rho \dd V &= \int_{\partial \Sigma} \dd F
\end{align}
where the 3D volume element is $\dd V = \sqrt{\gamma}~\dd x^3$, the density is $\alpha J^t$ and $\dd S_i$ is the vector surface element and $F$ is the flux across $\partial \Sigma$. If $\Sigma$ is a sphere of constant $R$ then $\dd S_i = \partial_i R$.
We can thus define
\begin{equation}
    \rho_E = -\alpha T^t_t, \quad \rho_J = \alpha T^t_{\phi}
\end{equation} 
where $\rho_E$ is the mass density and $\rho_J$ is the density of angular momentum about the BH axis. The rate of change in the total scalar field mass or total angular momentum between the BH horizon and an outer sphere is given by the integral of the respective flux across the sphere minus that across the horizon,  
\begin{equation}
    \partial_t M_{\textup{cloud}} = \int_{R=R_{\textup{max}}} \dd F  - \int_{R=R_+} \dd F.
    \label{eq:change_sphere_mass}
\end{equation}
In the limit $R_{\textup{max}} \rightarrow \infty$ the flux across the outer surface also corresponds to the change in the Arnowitt-Deser-Misner (ADM) mass or ADM angular momentum of the enclosed spacetime. They are therefore closely tied to physically measurable properties of the system, even at (large) finite distances, whereas the details of the distribution close to the BH are more observer dependent.
Further details are given in Appendix \ref{App:Flux}, where this is used to validate the code evolution. 


\subsection{Numerical Implementation}
We solve the second order Klein-Gordon equation by decomposing it into two coupled first order equations:
\begin{align}
\partial_t \varphi &= \alpha \Pi +\beta^i\partial_i \varphi \label{eqn:dtphi} ~ , \\ 
\partial_t \Pi &= \alpha \gamma^{ij}\partial_i\partial_j \varphi +\alpha\left(K\Pi -\gamma^{ij}\Gamma^k_{ij}\partial_k \varphi -\frac{dV(\varphi)}{d\varphi}\right)\nonumber  \\
& + \partial_i \varphi \partial^i \alpha + \beta^i\partial_i \Pi \label{eqn:dtPi} ~ ,
\end{align}
where $\Pi$ is the conjugate momentum density, as defined by Eqn. (\ref{eqn:dtphi}) and $K$ is the trace of the extrinsic curvature $K_{ij} = \frac{1}{2\alpha} \left( -\partial_t \gamma_{ij} + D_i\beta_j + D_j \beta_i \right)$.

We use an adapted version of the open source code $\textsc{GRChombo}$ \cite{GRChombo} to solve Eqs. (\ref{eqn:dtphi}) and (\ref{eqn:dtPi}) on a fixed metric background in the QIK coordinates described above. The scalar field is evolved by the method of lines with 4th order finite difference stencils, Runge Kutta time integration and a hierarchy of grids with 2:1 resolution. The value of the metric and its derivatives are calculated locally from the analytic expressions at each point. Details of code validation tests and convergence are provided in Appendix \ref{App:Flux}.

The size of the simulation domain is $L=1024M$, and we use seven (2:1) refinement levels with the coarsest having $128^3$ grid points, although we use the bitant symmetry of the problem in Cartesian coordinates to reduce the domain to $64 \times 128^2$ points. We implement non-zero, time oscillating boundary conditions for the scalar field by extrapolating the field linearly in the radial direction from values within the numerical domain.

The form of the metric naturally imposes ingoing boundary conditions at the horizon, due to the causal structure of the black hole. At spatial infinity we extrapolate the field value within the grid radially at first order, to simulate the effects of a roughly constant energy density. This in effect allows both ingoing and outgoing modes, but can introduce unphysical effects in very long simulations - these can be easily identified by varying the domain size, but ultimately limit the time for which the growth can be studied. The time before strong boundary effects occur is of the order of the light crossing time for our simulation box of $1024M$. This is roughly $5$ms for a solar mass BH and $60~$days for a SMBH of mass $10^9M_{\odot}$.

\section{Analytic framework}
\label{sec-analytic}
In this section we summarise what is known for stationary solutions, and develop several approximate analytic tools and a perturbative formalism to understand the growth of the scalar hair over time in different regimes of the parameter space. These methods are then confirmed within their regime of validity by the full numerical results in Sec. \ref{sec-numerics}.

\subsection{Stationary solutions}\label{sec:stationary_sols}

There is no simple, exact analytic solution for the growth of the scalar cloud from a general initial state. However exact analytic quasi-stationary oscillatory solutions have been found which we would expect to describe the final state of the cloud which forms \cite{Kerr_sols}. These solutions can generically be expressed in BL coordinates as:
\begin{equation}
    \varphi_{lm}(t, r,\theta,\phi) = \textup{Re}\left\{\Phi_0 e^{-i\omega t} R_{lm}(r) e^{im\phi}S_{lm}(\theta; ic)\right\},
    \label{Kerrsol}
\end{equation}
where $\Phi_0$ describes an initial amplitude. Here, $e^{im\phi}S_{lm}(\theta; ic)$ are oblate spheroidal harmonics, which reduce to spherical harmonics when $c = ak = 0$, where $k = \sqrt{\omega^2 - \mu^2}$ is the (complex) momentum at infinity. When $\omega$ is real, the solutions are stationary in time, whereas if $\omega$ has an imaginary component the field grows or decays exponentially with time. The radial functions $R_{lm}$ satisfy an equation of the form
\begin{equation}
    \left[\partial^2_{r_*} - V(r)\right]\sqrt{r^2 + a^2}R_{lm}(r) = 0,
    \label{eq:radial_eq}
\end{equation} 
where $r_*$ is the tortoise coordinate 
\begin{equation}
     r_* = r + \frac{2M}{r_+ - r_-}\left[r_+ \ln(r - r_+) - r_- \ln(r - r_-)\right].
     \label{eq:tortoise}
\end{equation}
The general solutions are given by:
\begin{equation}
    R_{\pm,lm}(r) = \; e^{\mp \tfrac{1}{2}\alpha z}z^{\pm \tfrac{1}{2}\beta}(z+1)^{\tfrac{1}{2}\gamma}\textup{HeunC}(\pm \alpha,\pm \beta,\gamma,\delta,\eta,-z),
\end{equation}
where $\pm$ denotes modes which are ingoing/outgoing at the horizon. Here,
HeunC is the confluent Heun function \cite{Heun} with 
\begin{align}
    z &= \frac{r - r_+}{r_+ - r_-}, \quad \alpha = 2i(r_+ - r_-)k,\quad \beta = 2i\frac{(am - 2r_+M\omega)}{r_+ - r_-},\nonumber \\
    \gamma &= 2i\frac{(am - 2r_-M\omega)}{r_+-r_-}, \quad \delta = -2M(r_+ - r_-)(\omega^2 + k^2), \nonumber\\
    \eta &= (\omega^2 + k^2) r^2_+ + \omega^2 (a^2 + 2M^2) - \frac{(am - 2M\omega)^2}{2(1 - a^2/M^2)} - \lambda_{lm},
\end{align}
where $\lambda_{lm}$ are the eigenvalues of the oblate spheroidal harmonics \cite{Kerr_sols}. In the limit $r \rightarrow r_+$ i.e.~$z \rightarrow 0$, we obtain   
\begin{equation}
    R_{\pm,lm}(r) \rightarrow \exp(\mp i k_H r_*), 
\end{equation}
up to a constant phase factor \cite{Superradiance_Review}, where $k_H = \omega - \frac{am}{2Mr_+}$. For the limit $r \rightarrow \infty$ there are also two independent ingoing/outgoing solutions which are independent of the spin of the black hole. In particular, for $k \neq 0$ they are
\begin{equation}
    R^{\infty}_{\pm,lm}(r) \approx \frac{e^{\mp ik(r-r_+)}}{r}z^{ \mp i\kappa}e^{\pm\pi\kappa},
\end{equation}
 where $\kappa = M(\omega^2 + k^2)/k$. For $\omega = \mu$ i.e. $k=0$
\begin{equation}
    R^{\infty}_{\pm,lm}(r) \approx r^{-3/4}e^{\mp i2\mu\sqrt{2Mr}}.
    \label{eq:far_field_sol}
\end{equation}
Note that in general each ingoing/outgoing solution at the horizon tends to a linear combination of the ingoing/outgoing solutions at $r \rightarrow \infty$ \cite{Berti_2009}. 

For non-spinning black holes, these analytical results have been shown to match well the spatial profile of the scalar cloud after long enough times \cite{Clough_2019}. For spinning black holes, we will see that the same holds.

As mentioned in the introduction, a well-known mechanism which enhances a scalar field around a spinning black hole is superradiance. In this case, the growth of hair is powered solely by the spindown of the black hole. It does not require the light boson to be dark matter, and indeed one assumes the field goes to zero far from the black hole \cite{Arvanitaki_2015}. Hence one rejects as unphysical any solutions \eqref{Kerrsol} which have energy ``entering from infinity" or ``escaping the black hole horizon", i.e.~one requires that solutions of the form \eqref{Kerrsol} to be \textit{ingoing} at the horizon and \textit{outgoing} at spatial infinity. These boundary conditions restrict $\omega$ to a discrete set of ``Quasi-Normal Modes" of the system \cite{Berti_2009}, indexed by overtone number $n = 0, 1, 2, ...$. The maximum size of the boson cloud is determined by the instability condition \eqref{eq:Sup} - as the black hole spins down $\Tilde{\Omega}_H$  decreases until \eqref{eq:Sup} becomes saturated \cite{Arvanitaki_2015}. Numerical simulations have shown that the maximum mass of cloud for a vector boson (Proca field) is $\sim 10\%$ the mass of the black hole \cite{East_2017} (see also \cite{Herdeiro_2017, Pani:2012bp, Dolan:2018dqv, Percival:2020skc} for semi-analytic studies), and similar magnitudes are expected for the scalar case. The accretion of compact boson stars onto BHs has also been proposed as a mechanism to enhance scalar clouds \cite{Sanchis-Gual:2020mzb, Clough:2018exo}.

In contrast, in the case we study, the cloud growth is powered by the reservoir of asymptotic scalar density. The field must still be ingoing at the horizon, but we allow both ingoing and outgoing modes at infinity. 

Neglecting the backreaction there is no way for non-interacting DM to radiate and lose energy, so we expect accretion to populate energy states with the same frequency $\omega$ as that of the accreting matter. In free Minkowski space, a uniform field oscillates with frequency $\mu$ and has energy $\mu$, so we expect the scalar matter to populate marginally bound states with $\omega \approx \mu$. Scalar matter bound within some gravitational potential in a galaxy will have energy $\omega \leq \mu$, while DM with some non-zero momentum at infinity will populate $\omega > \mu$ states.  

\subsection{Effective potential}

In order to examine the accretion of the scalar cloud, we seek solutions of Klein-Gordon equation in the Kerr metric which grow with time instead of just oscillating. It is convenient to recast the KG equation in a Schr{\"o}dinger-like equation, where one can interpret the solutions as a scattering off of a barrier of an effective potential. 

\begin{figure}
    \centering
    \includegraphics{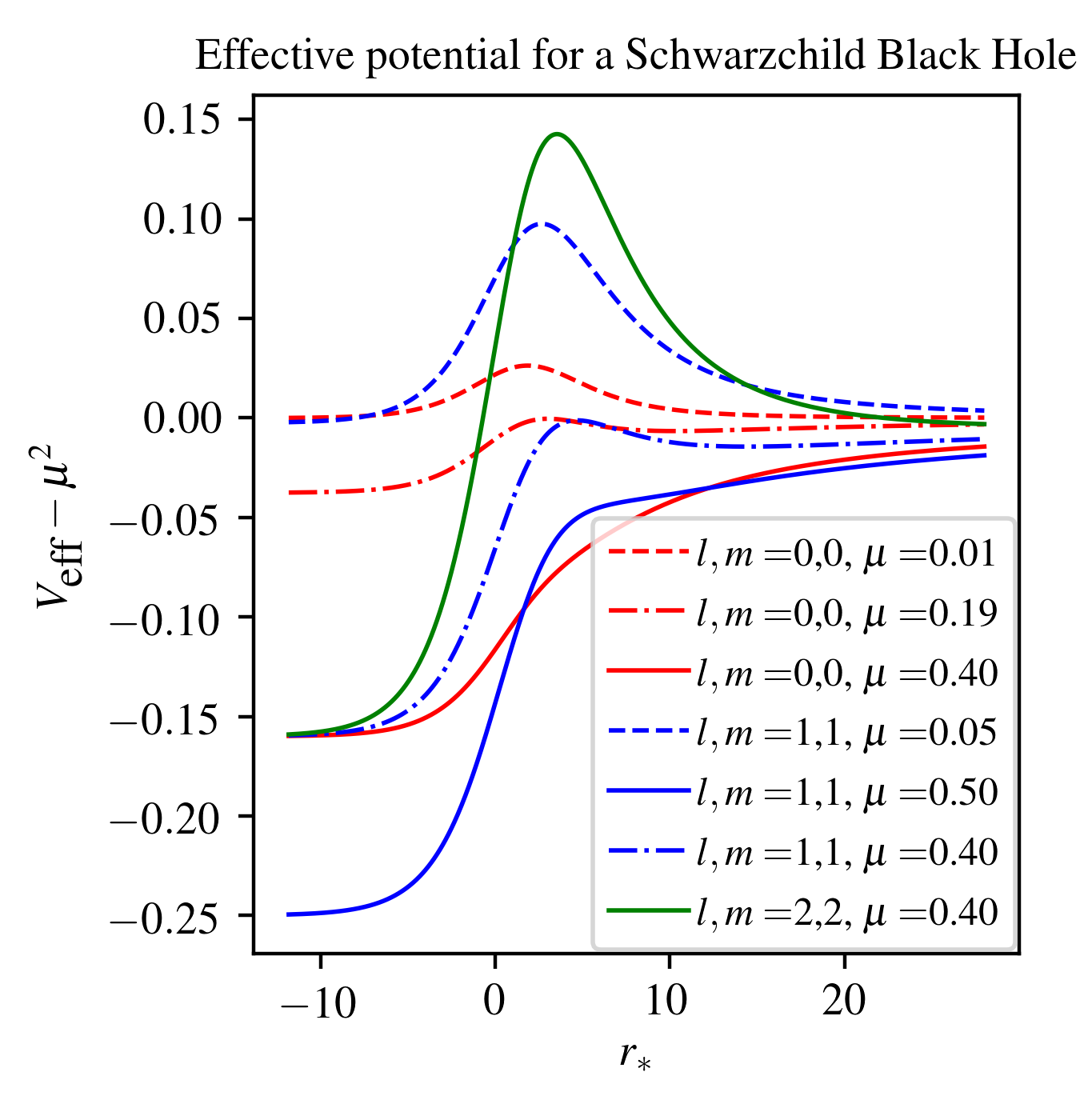}
    \caption{Schwarzschild effective potential \eqref{eq:Sch_eff_pot} for $M=1$ and different values of the scalar field mass and angular profile. As is well known, increasing the scalar angular momentum and lowering the scalar mass both tend to increase the potential barrier and thus allow stable clouds to form around the BH.}
    \label{fig:Sch_eff_pot}
\end{figure}

For simplicity, let us first consider a Schwarzschild background metric. In this case, the KG equation can be expressed as  
\begin{equation}
    \left[\partial^2_t - \partial^2_{r_*} + V_{\textup{eff}}(r; l, \mu) \right]\Psi_l(r, t) = 0,
    \label{eq:RW}
\end{equation} with 
\begin{equation}
\varphi_{lm} = \frac{1}{r}Y_{lm}(\theta,\phi)\Psi_l(r,t).
\end{equation}
Equation \eqref{eq:RW} takes the form of a Schr{\"o}dinger equation with effective potential 
\begin{equation}
    V_{\textup{eff}}(r;l, \mu) = \left(1 - \frac{2M}{r}\right)\left(\mu^2 + \frac{2M}{r^3} + \frac{\Lambda}{r^2}\right)
    \label{eq:Sch_eff_pot}
\end{equation}
where $\Lambda = l(l+1)$. Fig.~\ref{fig:Sch_eff_pot} shows potential profiles for different values of $M\mu$ or $l$.
We see that the potential barrier is lower for higher scalar masses, which allows the scalar matter to simply infall towards the black hole, whereas for lower masses the barrier is higher and reflects ingoing scalar waves, thus generating standing waves around the black hole. Physically, this is due to the pressure support generated by gradients of the scalar in the low mass case. These results have been studied analytically \cite{Hui_2019} and numerically \cite{Clough_2019} for $l=m=0$. In \cite{Clough_2019} and \cite{Hui_2019} only spherically symmetric profiles were considered, however here we extend our analysis to environments where the scalar field has asymptotic angular momentum. We see that the higher the $l,m$, and thus the higher the asymptotic scalar angular momentum, the higher the potential barrier. Hence we expect that increasing the angular momentum of the scalar cloud will reduce the accretion rate onto the BH. 

For the Kerr metric in Boyer-Linquist coordinates, the Klein-Gordon equation can be similarly written as
\begin{equation}
    \left[\partial^2_t - \partial^2_{r_*} + V_{\textup{eff}}(r; a, l, m,\mu) 
    + \hat{\mathcal{L}}_1\right]\Psi_{lm}(r, \theta, t) = 0,
    \label{eq:Kerr_KG}
\end{equation}
where  
\begin{equation}
\begin{split}
    \hat{\mathcal{L}}_1 =& -\frac{\Delta}{(a^2 + r^2)^2}\bigg[\frac{2\partial_{\theta}Y_{lm}}{Y_{lm}}\partial_{\theta} + \frac{\partial_{\theta}}{\sin\theta}\big(\sin\theta\partial_{\theta}\big)\\ &+ a^2\sin^2\theta(\partial^2_t + \mu^2) \bigg]
    + \frac{4 i m M a r}{(a^2 + r^2)^2}(\partial_t+i\mu),
\end{split}
\end{equation}
and 
\begin{equation}
    \varphi_{lm} = \frac{Y_{lm}(\theta, \phi)\Psi_{lm}(r, \theta, t)}{\sqrt{a^2 + r^2}}.
\end{equation}
For growing non-stationary solutions we cannot fully separate variables to simplify this equation. However, by writing \eqref{eq:Kerr_KG} in this form we can see that in the limit $a \rightarrow 0$, where $\Psi_{lm}(r, \theta, t) \rightarrow \Psi_{l}(r, t)$, we find $\hat{\mathcal{L}}_1 \Psi_{lm} \rightarrow \hat{\mathcal{L}}_1 \Psi_{l}(r, t) = 0$ as $\partial_{\theta} \Psi_{l}(r, t) = 0$. Hence the $\hat{\mathcal{L}}_1$ operator is capturing the deviation from a Schr{\"o}dinger-type equation due to the black hole's spin. 

\begin{figure}
    \centering
    \includegraphics{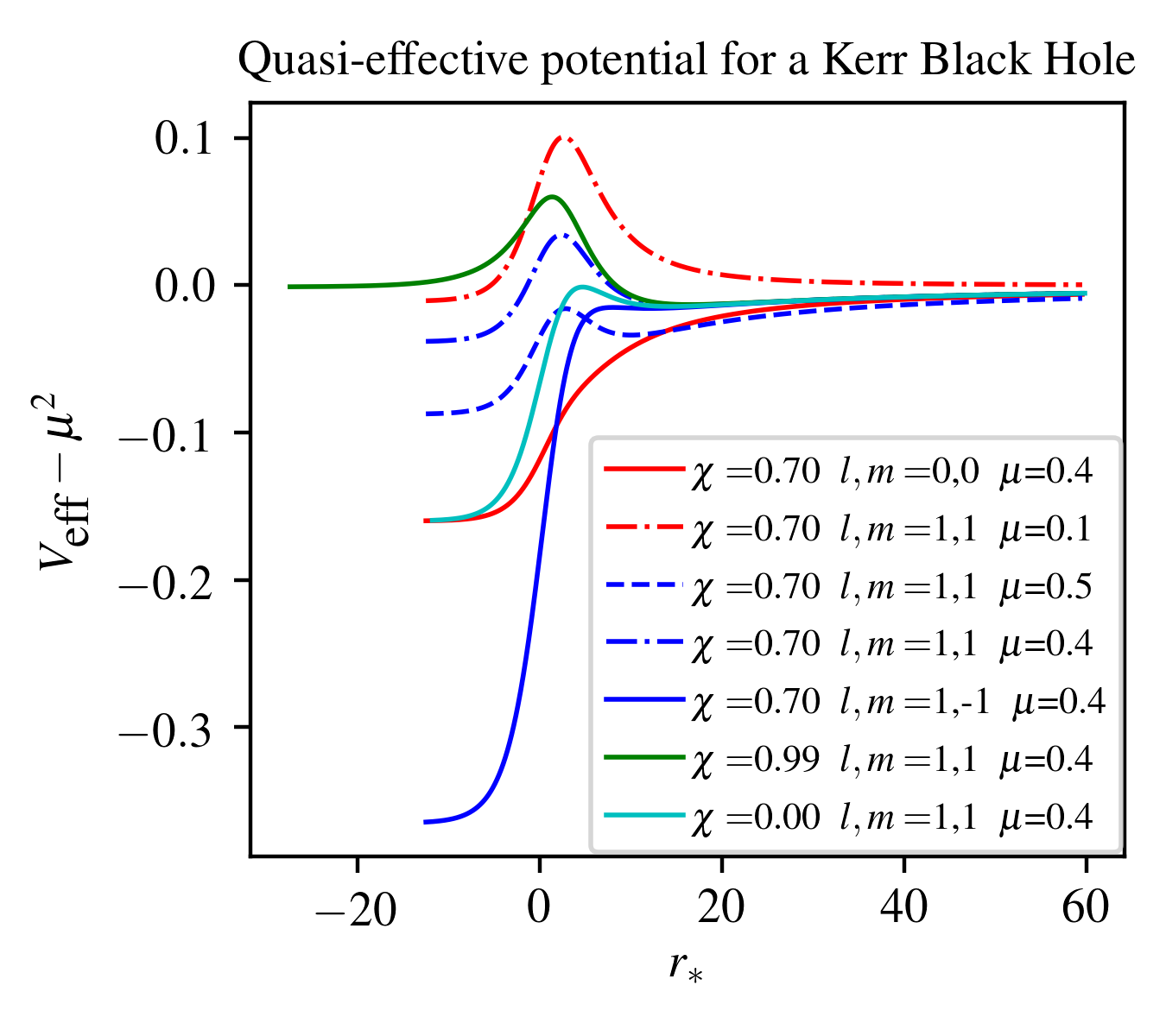}
    \caption{Kerr quasi-effective potential \eqref{eq:Kerr_eff_pot} for $M=1$ and different values of the scalar mass, angular profile, and BH spin. Similarly to Fig. \ref{fig:Sch_eff_pot}, we see that decreasing the scalar mass, adding a co-rotating angular momentum to the field and increasing the BH spin all tend to increase the potential barrier and therefore decrease accretion onto the BH. Note that for fixed mass and spin, the profiles are similar at larger distances, even in the co and counter rotating cases (green and blue solid lines).}
    \label{fig:Kerr_eff_pot}
\end{figure}

If we simply ignore the complications and mode-mixing induced by $\hat{\mathcal{L}}_1$, then $V_{\textup{eff}}$ is a quasi-effective potential given by
\begin{equation}
\begin{split}
    V_{\textup{eff}}(r) =& \frac{\Delta}{a^2 + r^2}\bigg[\mu^2 + \frac{a^2}{(a^2 + r^2)^2} + \frac{2Mr(r^2 - 2a^2)}{(a^2 + r^2)^3} \\
    &+ \frac{l(l+1)}{(a^2 + r^2)}\bigg] + \frac{am(4M\mu r - am)}{(a^2 + r^2)^2}.
    \label{eq:Kerr_eff_pot}
\end{split}
\end{equation}
which is the same as that given in Arvanitaki and Dubovsky (2011) equation (21) \cite{Arvanitaki_2011}.
We plot this quasi-effective potential in Fig.~\ref{fig:Kerr_eff_pot}, where we see that it behaves in a similar way to the Schwarzschild effective potential. Again we see that decreasing the scalar mass or increasing the $l,m$ number, and thus increasing the angular momentum, increases the potential barrier to infall. However we also see that changing the cloud and BH angular momentum from aligned ($l=m=1$) to anti-aligned ($l=1, m=-1$) eliminates the potential barrier, suggesting it would lead to faster accretion.
The intuition provided by these approximate ``potentials'' will be validated in our numerical studies.

\subsection{Particle limit}

\begin{figure}
    \centering
    \includegraphics{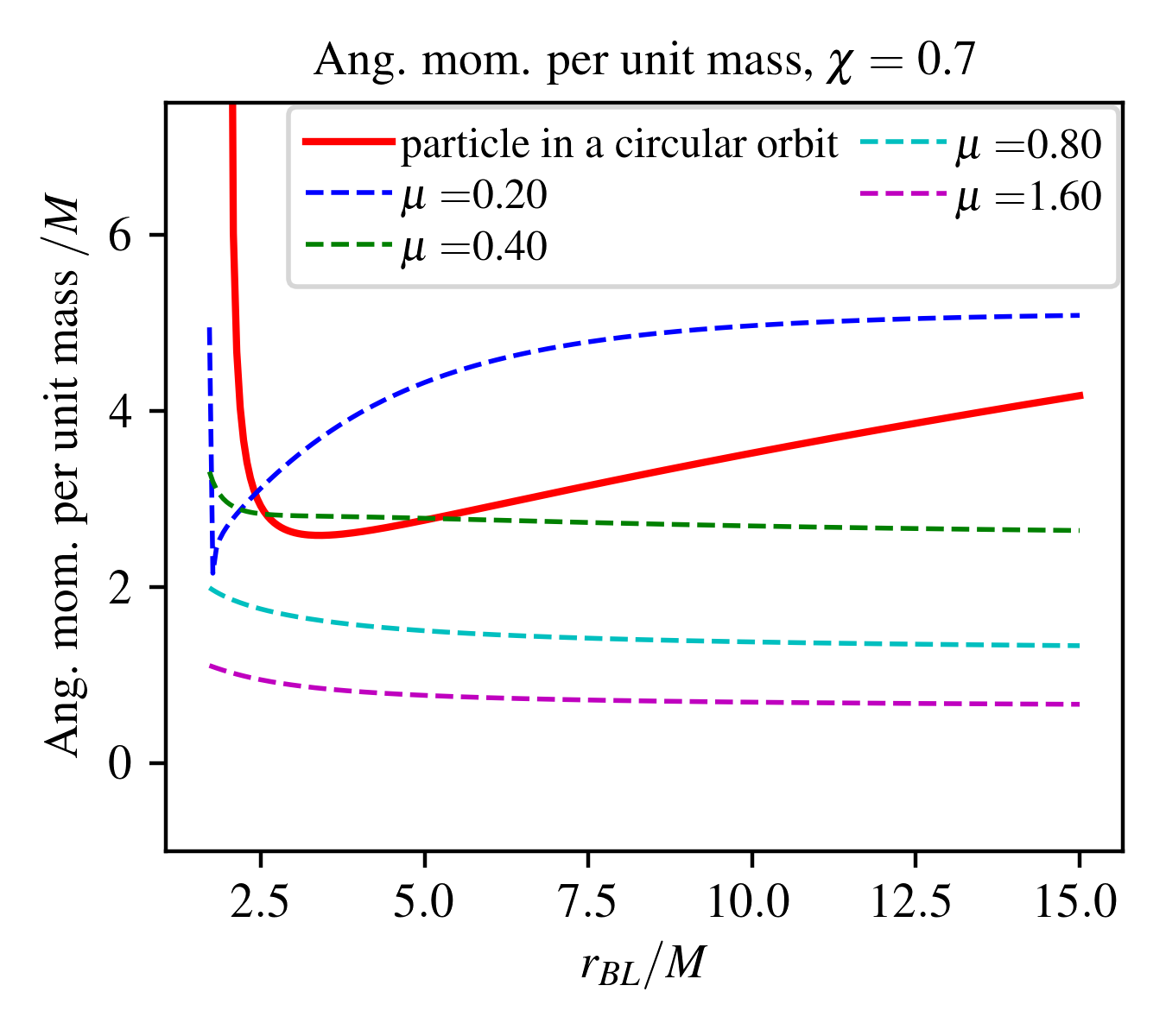}
    \caption{The solid red line shows the angular momentum per unit mass in the $z=0$ plane for a single massive particle in a circular orbit. The dashed lines show the angular momentum per unit mass of the scalar field at $t=0$ (given by equation \eqref{eq:field_Lz}) for different initial conditions, all with $l=m=1$ but with different $\mu$. If the dashed line lies above the red solid line (for a circular orbit) we expect the scalar field particles to move outwards, otherwise they will fall towards the black hole. Here we see the critical case around $\mu=0.4$ for which the cloud should concentrate close to the BH, whereas lower mass cases form clouds further out and higher mass cases accrete onto the BH.}  \label{fig:ang_mom_vs_particle}
\end{figure}

We can also make an analytical study of the scalar field behaviour in the particle limit. A massive particle in a circular orbit around a Kerr black hole in the $z=0$ plane has an angular momentum per unit mass given by \cite{KerrCircOrbit}
\begin{equation}
    h = \frac{r^2 + a^2 - 2a\sqrt{Mr}}{(r(r^2 - 3Mr + 2a\sqrt{Mr})/M)^{1/2}}.
    \label{eq:particle_Lz}
\end{equation}
Then for a complex scalar field (using the complex case to make the formula simpler) of the form $\varphi = e^{-i\mu t}Y^*_{lm}(\theta, \phi)$ we have 
\begin{equation}
    \rho_J/\rho_E = \frac{2m\left[2Ma(a\mu - m) + \mu r(a^2+r^2)\right]}{2\mu^2\left[M(a^2 - r^2) + r(a^2 + r^2)\right] + m^2(r - 2M)},
    \label{eq:field_Lz}
\end{equation}
in Boyer-Lindquist coordinates in the $z=0$ plane assuming $l+m$ is even. Strictly speaking we should only treat the scalar field with a particle rather than a wave description when $M\mu \gg 1$ so the Compton wavelength is much smaller than the size of the black hole. However this particle treatment may still provide a useful heuristic even at small $M\mu$. 
Fig.~\ref{fig:ang_mom_vs_particle} shows equations \eqref{eq:particle_Lz} and \eqref{eq:field_Lz} plotted for $m=1$, $\chi = 0.7$ and $M\mu = 0.2, 0.4, 0.8, 1.6$. If the angular momentum exceeds the expected particle angular momentum per unit mass for a circular orbit we expect the equivalent particles will move outwards. If the angular momentum is less than expected for a circular orbit we expect they will fall towards the black hole. 

\subsection{Perturbative solutions for cloud growth}\label{sec:perturbative_analytic}
The growth of the scalar cloud can also be studied analytically with a perturbative approach for large $r$.
We start by rewriting the Klein-Gordon equation as 
\begin{equation}
    \mathcal{L}_{KG}\varphi = 0;\quad \mathcal{L}_{KG} = \frac{-1}{\sqrt{-g}}\partial_{\mu}\left(g^{\mu\nu}\sqrt{-g}\partial_{\nu}\right) + \mu^2.
\end{equation}
and choose units such that $M=1$. For a Kerr metric in Boyer-Linquist coordinates we can multiply out the denominator and expand in powers of $1/r$
\begin{align}
    \mathcal{L} = \frac{\Delta \Sigma}{r^4}\mathcal{L}_{KG}= \sum^6_{n=0}\frac{1}{r^n}\mathcal{L}_n, 
\end{align}
where 
\begin{align}
    \mathcal{L}_0 &= -(\partial^2_t + \mu^2), \\
    \mathcal{L}_1 &= \; 2\mu^2, \\
    \mathcal{L}_2 &= \; r^2 \partial^2_r + 2r\partial_r - \hat{L}^2 + a^2(1 + \cos^2\theta)\mathcal{L}_0, \\
    \mathcal{L}_3 &= -2\left(2r^2\partial^2_r +3r\partial_r - \hat{L}^2 - a^2\mu^2 + 2a\partial_t\partial_{\phi} - a^2\sin^2\theta \mathcal{L}_0\right), \\
    \mathcal{L}_4 &= \; 2(a^2 + 2)(r^2\partial^2_r + r\partial_r) + a^4\cos^2\theta \mathcal{L}_0 - a^2\partial^2_{\phi} - a^2\hat{L}^2, \\
    \mathcal{L}_5 &= -2a^2(r\partial_r + 2r^2\partial^2_r), \\
    \mathcal{L}_6 &= \; a^4 r^2 \partial^2_r,
\end{align}
and $\hat{L}^2$ is the spherical harmonic operator 
\begin{equation}
    \hat{L}^2 := -\frac{\partial_\theta}{\sin\theta}\left(\sin\theta \partial_{\theta}\right) - \frac{\partial^2_{\phi}}{\sin^2\theta}.
\end{equation}
Similarly, we expand the scalar field for large $r$ as:
\begin{equation}
    \varphi = \sum^\infty_{n=0} \frac{1}{r^n}\varphi_n. 
\end{equation}
We choose a (complex) $\varphi_0$ such that it satisfies our initial conditions and $\mathcal{L}_0 \varphi_0 = 0$
\begin{equation}
    \varphi_0 = \Phi_0 e^{-i\mu t}Y^*_{lm}(\theta, \phi),
\end{equation}
where $\Phi_0$ is a constant amplitude. Then we iteratively compute $\varphi_n$ by matching powers of $1/r$:
\begin{equation}
    \varphi_n = \mathcal{L}^{-1}_0\left(\sum^{n-1}_{j=0}\mathcal{L}_{n-j}\varphi_j\right).
\end{equation}
where the operator $\mathcal{L}^{-1}_0$ acting on a general function $A(t)$ behaves as:
\begin{equation}
    \mathcal{L}^{-1}_0 A(t) = e^{-i\mu t}\int^t_0 e^{2i\mu t_1}\int^{t_1}_0 e^{-2i\mu t_2} A(t_2) \dd t_2 \dd t_1. 
\end{equation}
Up to order $r^{-2}$ we obtain for the complex scalar field
\begin{equation}
\begin{split}
    \varphi(t) &\approx \varphi_0(t,\theta,\phi) \left[1+
    \frac{2 i \tau-e^{2 i
   \tau}+1}{2 \Tilde{r}}\right. \\
   &\left.-\frac{2\tau^2+2i\tau\left(\Tilde{\Lambda}-2-e^{2i\tau}\right)
   +(\Tilde{\Lambda}-3)\left(1-e^{2i\tau}\right)}{4\Tilde{r}^2}\right],
   \label{phi_sol}
\end{split}
\end{equation}
where $\Tilde{\Lambda}=l(l+1)/(M\mu)^2$, $\tau=\mu t$ and $\Tilde{r}=r/M$, putting back in the factors of $M$. Finally, we obtain a solution for a real scalar field $\varphi$ by taking the real part of \eqref{phi_sol}. Note that there is no dependence on BH spin at order $\Tilde{r}^{-2}$. This approximation is valid for $\Tilde{r}>\tau$, and therefore it only describes the initial growth of the cloud far from the black hole, and does not describe the late stationary state of the cloud where we expect $\varphi\propto r^{-3/4}$ for $M\mu \sim 1$ \cite{Hui_2019}.

Now recall from section \ref{sec:diagnostics} the flux element into a sphere is given by $\dd F = -\sqrt{-g}J^R$. In QIK coordinates metric determinant $\sqrt{-g}$ goes to zero on the horizon. For the stationary solutions $J^R$ diverges at the horizon in such a way that $\sqrt{-g}J^R$ is non-zero, however for the growing case $J^R$ is initially zero and remains finite at finite $t$. This means that for finite time the flux into the horizon is zero. The energy flux into a sphere of constant $R$ in QIK coordinates in time $t$ is then given by 
\begin{align}
    \partial_t M_{\textup{cloud}} = \int_{R=R_{\textup{max}}} \dd F &= \iint \Sigma \left(\pdv{r}{R}\right)\;T^R_t\;\sin\theta ~\dd \theta \dd \phi, \nonumber\\
    &= \iint \Sigma \;g^{rr} \partial_t\varphi \partial_r\varphi\;\sin\theta ~\dd \theta \dd \phi, \nonumber\\
    &= \iint \Delta \partial_t\varphi \partial_r\varphi\;\sin\theta ~\dd \theta \dd \phi, \nonumber\\
    &:= 2\pi \int \mathfrak{J}_r \sin\theta ~\dd \theta,
\end{align}
where $\mathfrak{J}^{t}_r := \frac{1}{2\pi} \int \Delta \partial_{t}\varphi \partial_r\varphi \dd \phi$. Hence the change in the total mass of scalar field inside radius $R$ but outside the horizon is given by 
\begin{equation}
    \delta M_{\textup{cloud}} = 2\pi \int \int^t_0 \mathfrak{J}_r\dd t' \sin\theta \dd \theta.
\end{equation}

Neglecting the oscillatory terms, we find the energy flux for large $\Tilde{r}$ to be described by
\begin{equation}
\begin{split}
\int^t_0 \mathfrak{J}_r \dd t' &= \Phi^2_0 \vert Y_{lm}\vert^2 \frac{1}{4}\bigg\{\tau^2 - \frac{(1+\Tilde{\Lambda})\tau^2}{\Tilde{r}} +\\ 
&\frac{\tau^2}{2\Tilde{r}^2}\bigg[7\Tilde{\Lambda} -\chi^2(3\cos2\theta+7)-12\Tilde{m}\bigg]+\\
&\frac{\tau^2}{2\Tilde{r}^3}\bigg[(\tfrac{4}{3}-\Tilde{\Lambda})\tau^2+ 2\chi^2(5\cos2\theta-\Tilde{\Lambda}+16) + \\
&40\Tilde{m} - \Tilde{\Lambda}^2-4\Tilde{\Lambda}-20\bigg]+\frac{\tau^2}{4\Tilde{r}^4}\bigg[\tfrac{1}{9}\tau^4+\\
&\tfrac{1}{6}\Big(3\chi^2(\cos2\theta + 3)+3\Tilde{\Lambda}^2+35\Tilde{\Lambda}-59+12\Tilde{m}\Big)+\\
& (6\chi^2+5\Tilde{\Lambda}+9)(\chi^2\cos2\theta+4\Tilde{m})+18\chi^4+\\
& \chi ^2(9\Tilde{\Lambda}+19)-6\Tilde{\Lambda}^2+\Tilde{\Lambda}-35)\bigg]+\mathcal{O}(r^{-5})\bigg\},
\label{eq:mass_flux_perturb}
\end{split}
\end{equation}
where $\Tilde{m}=\chi m/(M\mu)$.

We can do the same for the angular momentum flux. For $\mathfrak{J}^{\phi}_r := -\frac{1}{2\pi} \int \Delta \partial_{\phi}\varphi \partial_r\varphi \dd \phi$ we obtain 
\begin{equation}
    \begin{split}
        \int^t_0 \mathfrak{J}^{\phi}_r \dd t' &= \vert Y_{lm} \vert^2 \frac{M m}{4}\bigg\{\tau^2 - \frac{\Tilde{\Lambda}\tau^2}{\Tilde{r}} +\\ 
&\frac{\tau^2}{2\Tilde{r}^2}\bigg[4\Tilde{\Lambda} -a^2(3\cos2\theta+7)-12\Tilde{m}\bigg]+ \\
&\frac{\tau^2}{2\Tilde{r}^3}\bigg[(\tfrac{4}{3}-\Tilde{\Lambda})\tau^2+ 2a^2(3\cos2\theta-\Tilde{\Lambda}+11) + \\
&24\Tilde{m}+2\Tilde{\Lambda}-20\bigg]+
\dots\bigg\} .
    \end{split}
\end{equation}
again neglecting the oscillating terms.

From this result we see that the primary timescale for mass growth is the oscillation period $\mu^{-1}$ and that the most important factor affecting the growth rate is the quantity $\Tilde{\Lambda}$ which roughly corresponds to the angular momentum per unit mass squared. The BH spin only enters at order $1/r^2$, while the azimuthal number $m$ appears at lowest order as $-6\Tilde{m}\frac{\tau^2}{\Tilde{r}^2}$. This suggests that anti-aligned BH and cloud spins with $\Tilde{m} < 0$ gives faster growth than aligned spins. We can also note that the expression for the angular momentum flux is very similar to the expression for the mass flux, suggesting the angular momentum per unit mass is approximately constant in the regime of validity for this result (i.e., small $t$ and large $R$). 

It is also worth comparing to the expected flux for the stationary solutions in section \ref{sec:stationary_sols}. At large $r \gg M$ we find the ingoing solution to be 
\begin{equation}
    \varphi \approx \frac{\sqrt{\rho_{R_c}}}{\mu}\left(\frac{r}{R_c}\right)^{-3/4}\vert Y_{lm} \vert \cos(\mu t - m\phi + 2\mu\sqrt{2Mr}),
    \label{eq:explicit_far_field}
\end{equation}
where $\rho_{R_c}$ is the scalar field density at radius $R_c \gg M$. Hence 
\begin{equation}
    \mathfrak{J}^{\textup{stationary}}_r = \tfrac{1}{2\pi}\int \Delta \partial_t\varphi \partial_r\varphi \dd \phi \approx \rho_{R_c} \vert Y_{lm} \vert^2 R^{3/2}_c \sqrt{2M}, 
\end{equation}
which is independent of $r$ as required for a stationary solution. If we match the asymptotic scalar field density $\rho_0 = \frac{1}{2}\Phi^2_0\mu^2$ to $\rho_{R_c}$ then the perturbative solution \eqref{eq:mass_flux_perturb} is 
\begin{equation}
    \mathfrak{J}^{\textup{perturbative}}_r = \rho_{R_c}\vert Y_{lm} \vert^2 \left\{t - \frac{(1+\Tilde{\Lambda})t}{\Tilde{r}}+\dots\right\}.
\end{equation}
We can then intuit that if we start with homogenous conditions the scalar flux will grow from zero until it matches the stationary value.  

\section{Numerical Results}
\label{sec-numerics}

In this section we summarise our numerical results for different values of the parameters described in Sec. \ref{sec-methods}. Plots display results in the time parameter $\tau$, a dimensionless ratio of the physical time to the scalar field oscillation period, to enable clearer contact with the perturbative result. The physical time $t$ can simply be recovered as $t = \tau / \mu$ (ie, at some $\tau$, the physical time that has passed $t$ is longer for smaller $\mu$).

Where we show the profiles for the energy density as a function of radius, at each radius the density is averaged over a sphere of constant $r$ to remove the angular dependence, and normalise against $\rho_0 = \frac{1}{2}\Phi^2_0 \mu^2$, the asymptotic energy density.

\subsection{Adding spin to the black hole, zero scalar angular momentum}

\begin{figure}
    \centering
    \includegraphics{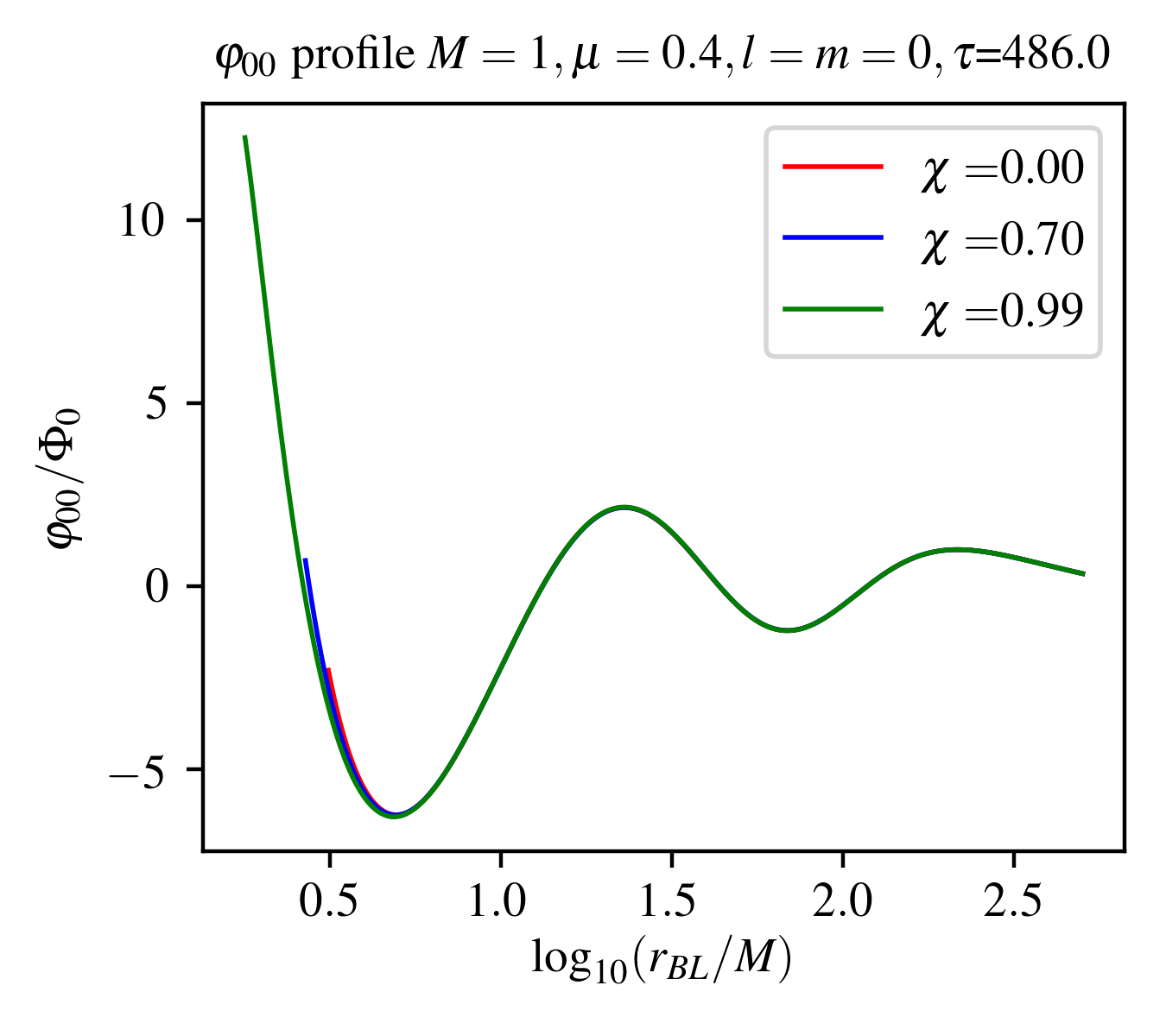}
    \caption{Scalar field $\varphi$ $l=m=0$ component as function of radius, for initial spherically symmetric profile $l=m=0$ in BL coordinates. Different BH spins are shown in different colours.
    We see that BH spin leads to very mild changes in the scalar cloud close to the horizon, but no significant change in the profile. 
    }
    \label{fig:phi_vs_a}
\end{figure}

\begin{figure}
    \centering
    \includegraphics{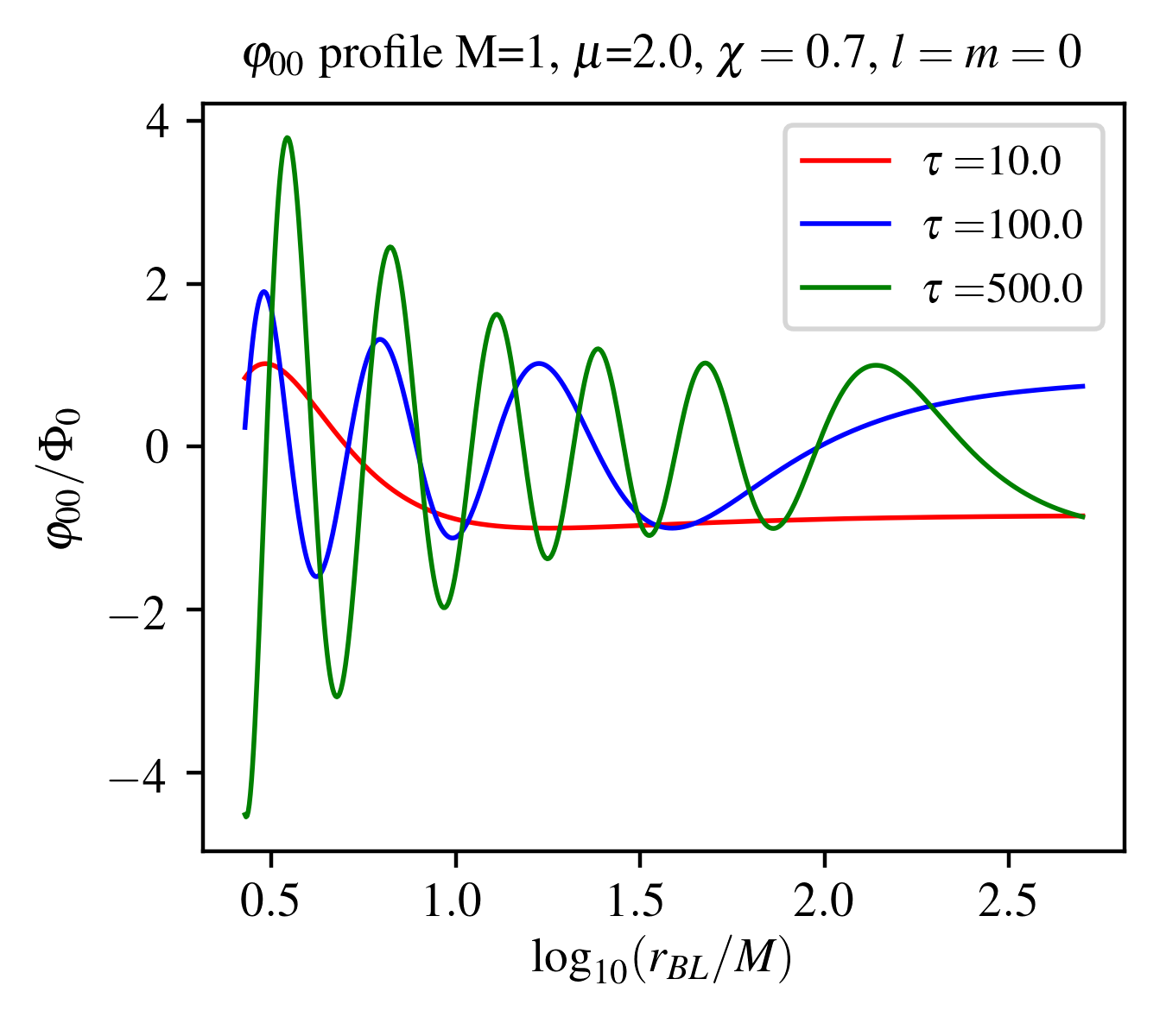}
    \caption{Radial $\varphi$ $l=m=0$ mode for $M\mu=2.0, \chi=0.7$ for different times in BL coordinates. We see oscillations spreading out from the horizon such that the $\varphi$ profile gradually converges towards the stationary solution.}
    \label{fig:phi_vs_t}
\end{figure}

\begin{figure}
    \centering
    \includegraphics{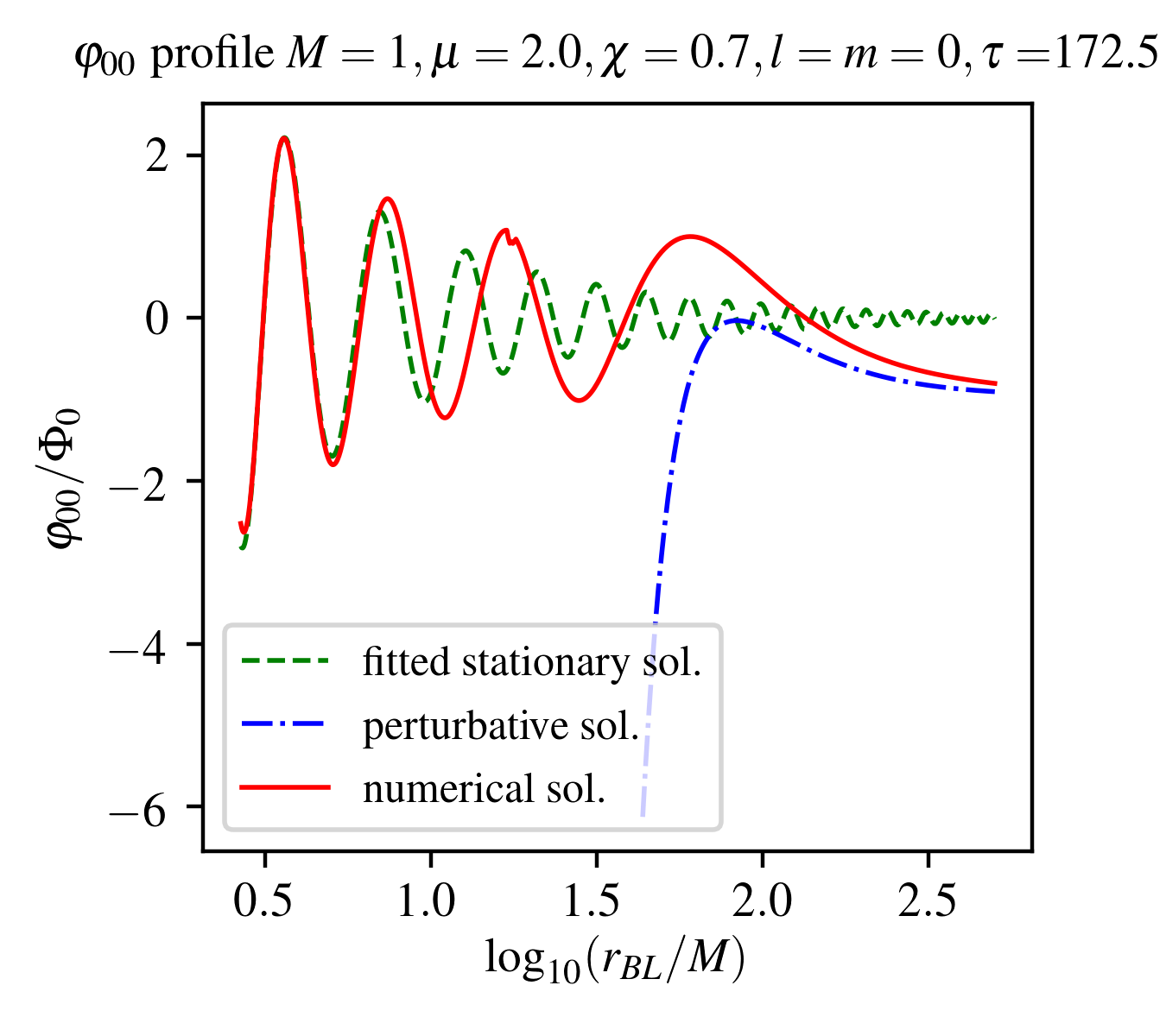}
    \caption{Radial $\varphi$ $l=m=0$ mode for $M\mu=2.0, \chi=0.7$ and $\tau=172.5$ in BL coordinates. We see that the stationary analytic solution is a good fit for the region close to the horizon, while the perturbative solution is a good fit for large $r$ where $r \gtrsim \tau$.}
    \label{fig:phi_vs_analitic}
\end{figure}

We can first examine the effect of adding BH spin to the Schwarzschild case examined in \cite{Clough_2019} - that is, where there is no angular momentum in the scalar. We find that, contrary to what one might expect, adding BH spin introduces only a very mild distortion of the field profile. In Fig.~\ref{fig:phi_vs_a} we show the radial profile of the $l=m=0$ component of a scalar field with $\mu=0.4$, in the case when the initial field was in the $l=m=0$ mode and thus spherically symmetric. We found similar results for the $l\neq 0$ components, with no observable excitement above the level of simulation error in the spinning cases.

In Fig.~\ref{fig:phi_vs_t} we show the radial profile of a scalar field with $\mu=2$ near the horizon, for a BH with dimensionless spin $\chi=0.7$, at several different times in the evolution. As in \cite{Clough_2019} we see that from the initially flat profile, the scalar develops radial oscillations first near the BH, which then develop outwards radially over time. The field profile oscillates in time and space with frequency/wavelength set by $\mu$. 

Finally, in Fig.~\ref{fig:phi_vs_analitic} we show the radial profile of a scalar field with $\mu=2.0$, for a BH with dimensionless spin $\chi=0.7$, and we compare the numerical with the analytical stationary and perturbative solutions discussed in Section \ref{sec:stationary_sols} and \ref{sec:perturbative_analytic}, respectively. We see that the stationary solution describes well the scalar field near the horizon, whereas the perturbative analytic expression describes the evolution far from the horizon, where the oscillatory behaviour has not been reached yet. The true solution interpolates between the two regimes.

\subsection{Adding angular momentum to the scalar, non zero black hole spin}

\begin{figure}
    \centering
    \includegraphics{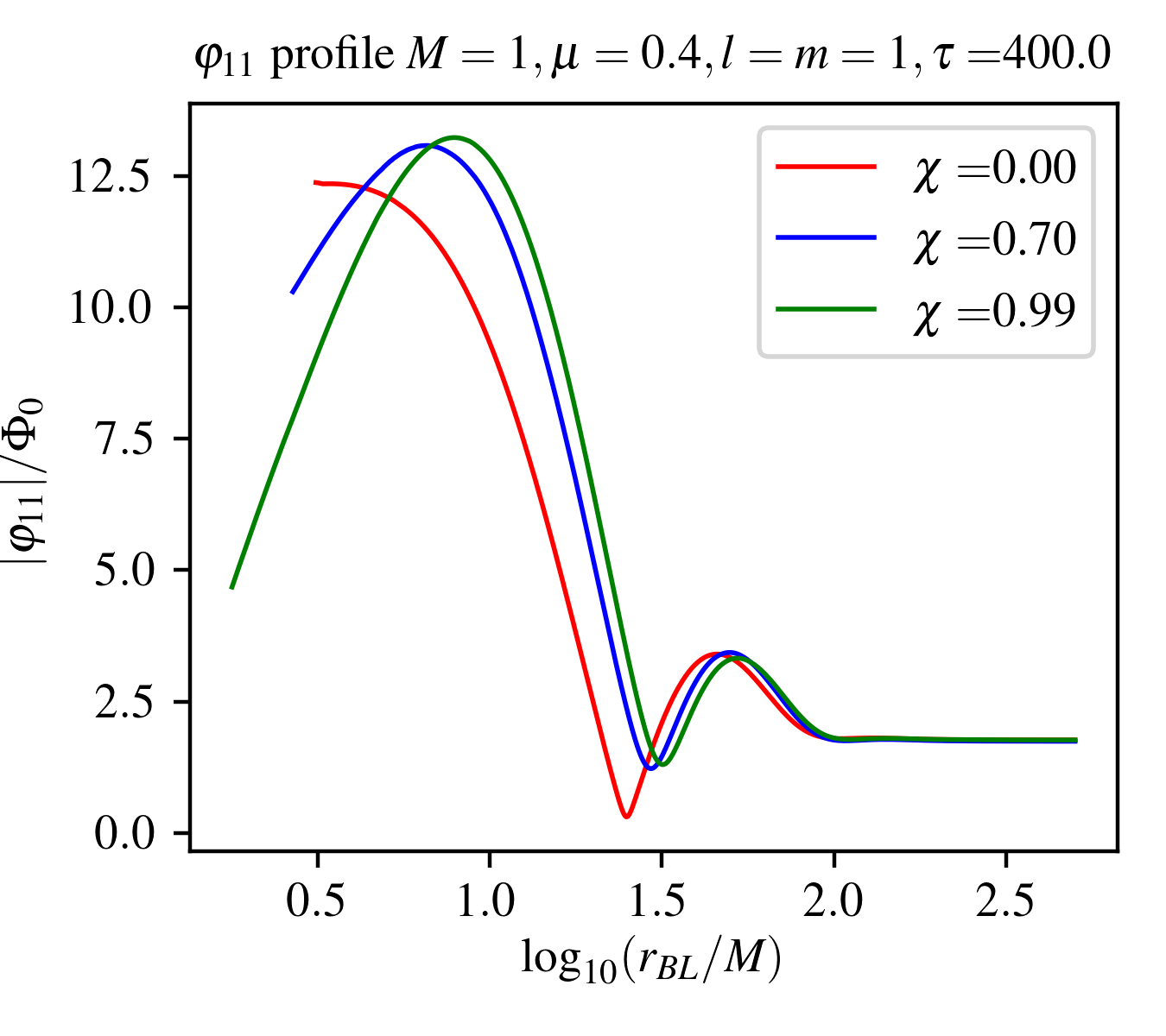}
    \caption{Scalar field $\varphi$ $l=m=1$ spherical harmonic component as function of radius for an initial $l=m=1$ angular dependence in BL coordinates. Different BH spins are shown in different colours as indicated. 
    BH spin again leads to changes in the scalar cloud close to the horizon which are more pronounced than in the $l=m=0$ case.}
    \label{fig:phi_vs_a_11}
\end{figure}

\begin{figure}
    \centering
    \includegraphics{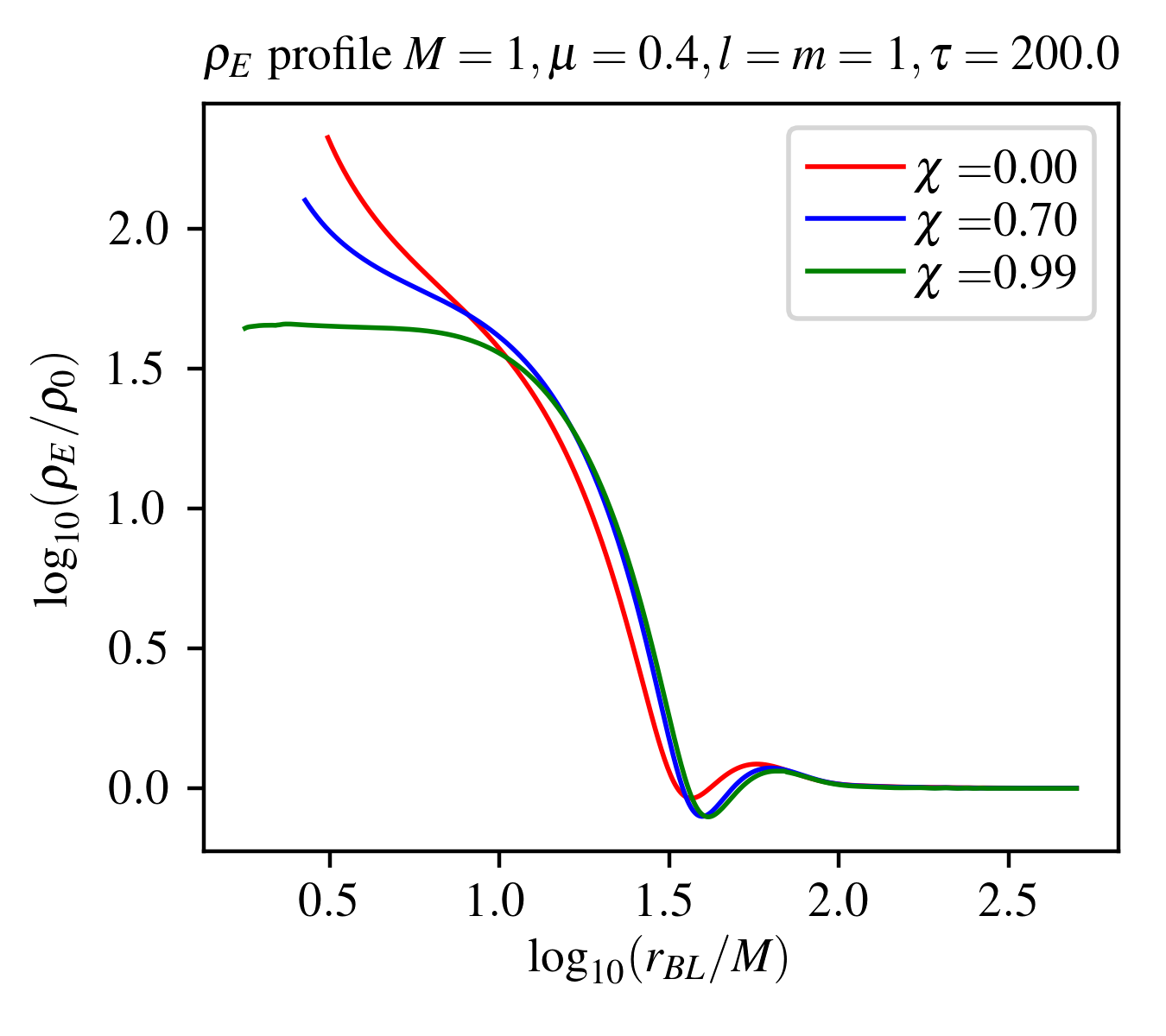}
    \caption{Radial energy density profile $\rho_E$ averaged over a sphere for $M=1,\mu=0.4,l=m=1$ and different $\chi$. Adding spin to the BH decreases the maximum energy density.}  \label{fig:rho_profile_compare_a}
\end{figure}

Next, we explore the impact of adding asymptotic angular momentum to the initial scalar field by choosing non-zero $l,m$ spherical harmonic numbers. A typical 2D profile was shown in QIK coordinates in Fig.~\ref{fig:spiral}. 

\subsubsection{Effect of BH spin $\chi$}
In Fig.~\ref{fig:phi_vs_a_11} we show the radial profile of the scalar field for simulations where the initial angular dependence was set by the $l=m=1$ spherical harmonic. Similarly to the previous section, we only plot the component of the scalar field mode with the same $l,m$, as we find that the other multipoles have negligible amplitude at all times. As in Fig.~\ref{fig:phi_vs_a}, we have $M\mu=0.4$ and different values of BH spin are shown in different colours. Changing the BH spin again modifies the profile close to the horizon, and the effect is much larger than for the $l=m=0$ case of Fig.~\ref{fig:phi_vs_a_11}. 
We see that adding spin to the BH decreases the maximum energy density, as shown in Fig \ref{fig:rho_profile_compare_a}.
However, further out there is little difference in the profiles, so as in the $l=m=0$ case we see minimal impact on the flux at a large radius.

\subsubsection{Effect of scalar angular momentum $l,m$}

\begin{figure}[h!]
    \centering
    \includegraphics{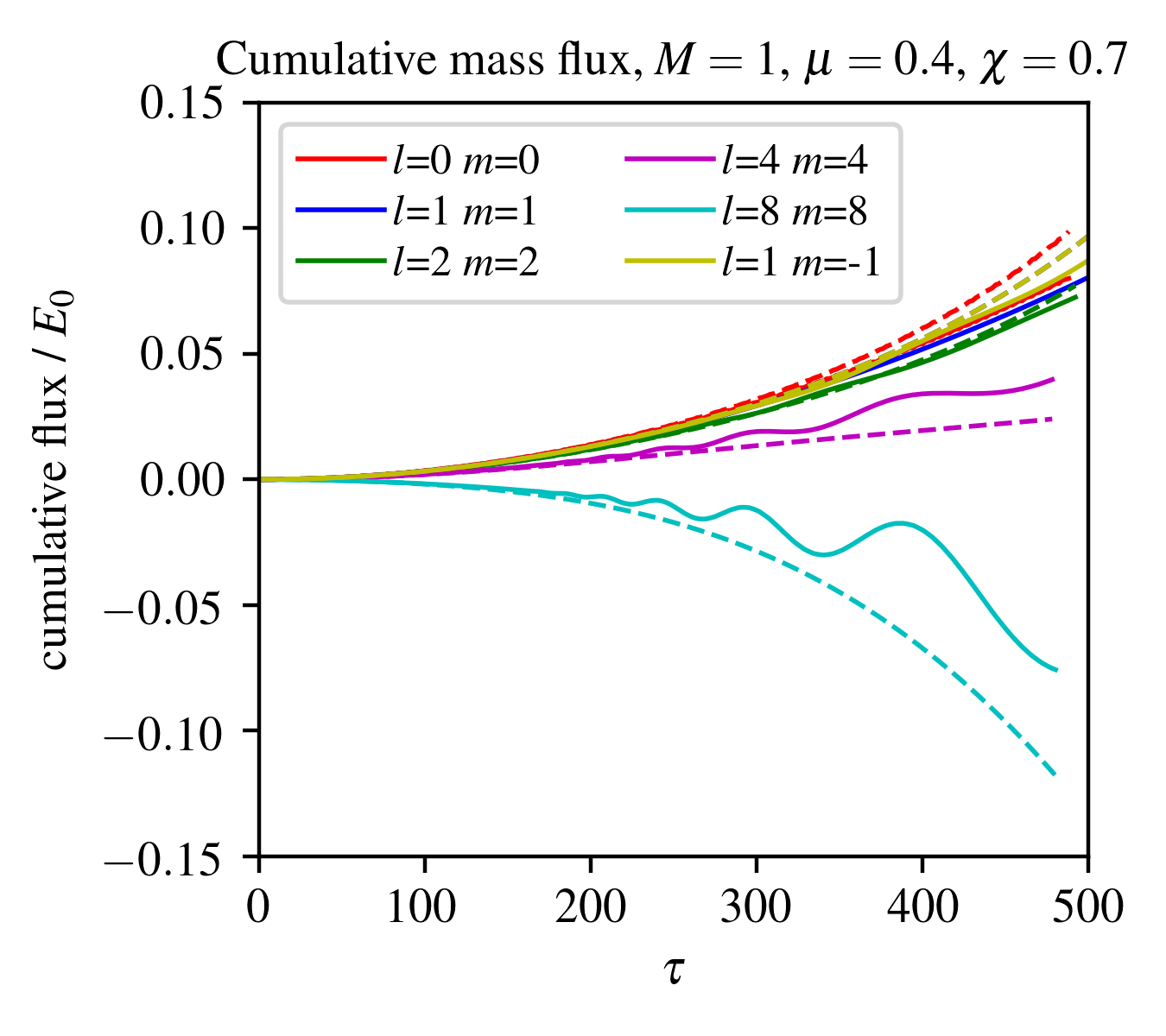}
    \caption{Cumulative flux of mass into a sphere with $R_E=300M$ around a Kerr BH for $\chi=0.7, \mu=0.4$ and different $l,m$. We normalise the flux by the quantity $E_0$ - the energy in a sphere of radius $R_E$ with constant energy density $\tfrac{1}{2}\mu^2\varphi^2_0$. Dashed lines describe the perturbative analytic flux to order $r^{-4}$, and the solid lines the numerical results. We see that accretion into the sphere is reduced for higher $l,m$ modes, with the negative flux in the highest case signalling that the cloud is forming outside $R_E$.}
    \label{fig:compare_lm}
\end{figure}

\begin{figure}[h!]
    \centering
    \includegraphics{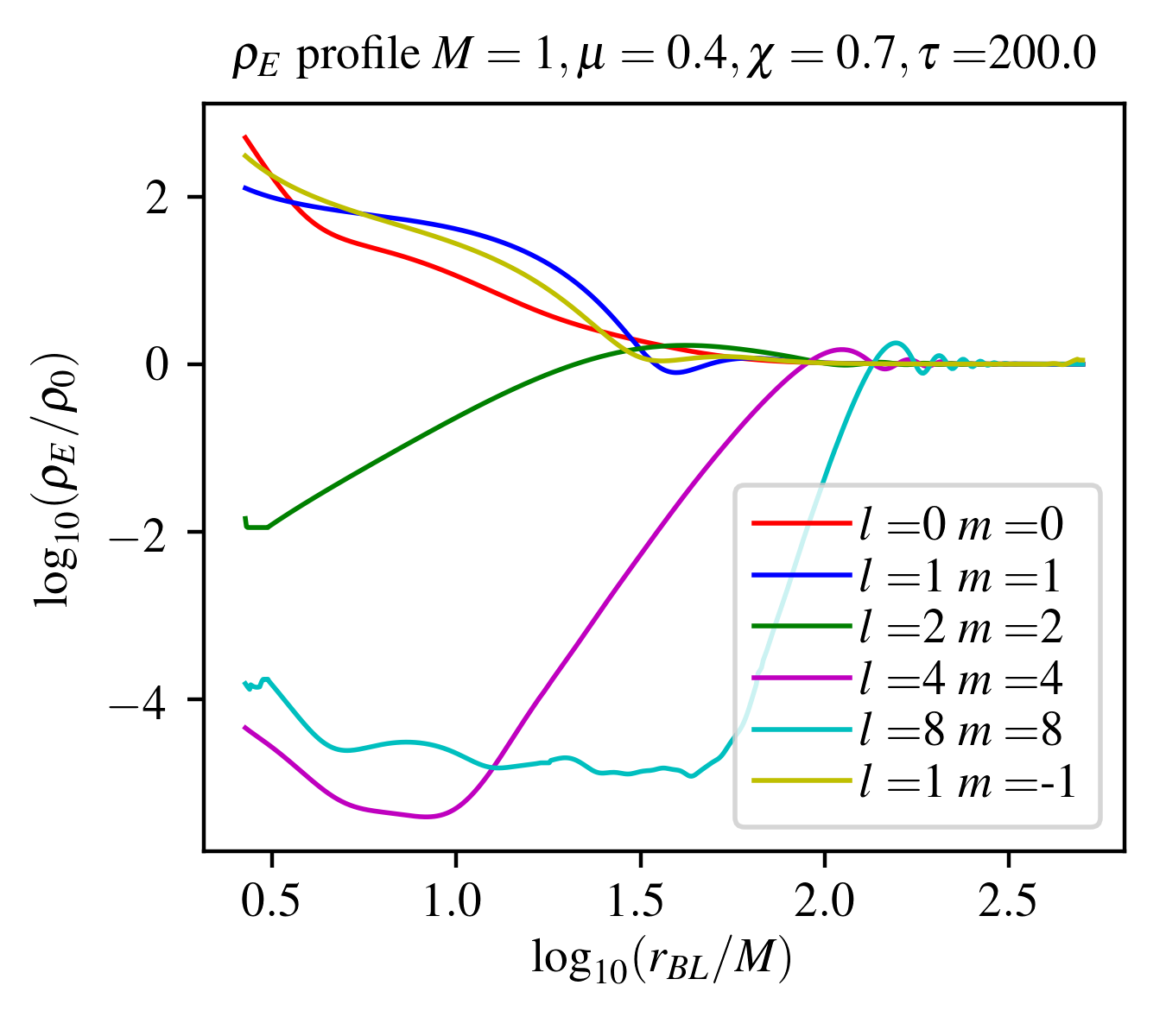}
    \caption{Radial profile of the $0,0$ harmonic mode of the energy density $\rho_E$, i.e. the energy density is averaged over a sphere at each radius. The parameters are $M\mu=0.4, \chi=0.7$ and the initial scalar field is set to different $l,m$ spherical harmonics at $t=0$.}  \label{fig:rho_profile_compare_lm}
\end{figure}

To quantify the growth rate of the cloud we extract the cumulative conserved scalar field energy flux through a sphere at a radius $R_E \gg M$. This equals the change in total conserved scalar field energy between the BH horizon and the sphere, and at large $R$ roughly corresponds to the change in the ADM mass of the spacetime due to the accretion. First, we show the cumulative flux for different $l,m$ modes in Fig.~\ref{fig:compare_lm}, along with the analytic perturbative expression that we derived in Sec. \ref{sec-analytic} above, to fourth order in $M/r$. We show results versus $\tau = \mu t$ as discussed above. The other important timescale is the free fall timescale set by the radius of extraction $R_E$. Radial oscillations in the profile first form near the BH, and gradually spread outwards on roughly this timescale, so we expect the stationary behaviour to be reached when these waves hit the extraction radius at approximately $t \propto R_E^{3/2}$, which is greater than the time period studied.

We see that the numerical result agrees well with the a perturbative expression for $\tau \ll r/M$ as we would expect. As $\tau$ increases the numerical result deviates from the analytic expression, with large $l,m$ producing the largest deviation. The dominant effect is from the first order $\delta M_c \sim \tau^2(1 - (1 + \Tilde{\Lambda})/\Tilde{r})$ term. Larger $l$ for fixed $\mu$ corresponds to larger $\Tilde{\Lambda}$ and thus larger cloud asymptotic angular momentum per unit mass, which increases the potential barrier to accretion, decreasing the growth rate. This is also consistent with what we saw from the effective potential and particle pictures, and physically corresponds to the fact that the cloud is forming further out from the BH - in the case of $l=m=8$ in the figure, this is even outside the extraction sphere, hence the overall decrease in the mass. Whilst this means that the cloud is not accreted onto the BH, it generally decreases the maximum energy density in the spacetime comapred to the accreting cases, due to it being spread out over a larger volume, as shown in Fig. \ref{fig:rho_profile_compare_lm}.

\subsubsection{Effect of scalar mass $\mu$}

\begin{figure}[h!]
    \centering
    \includegraphics{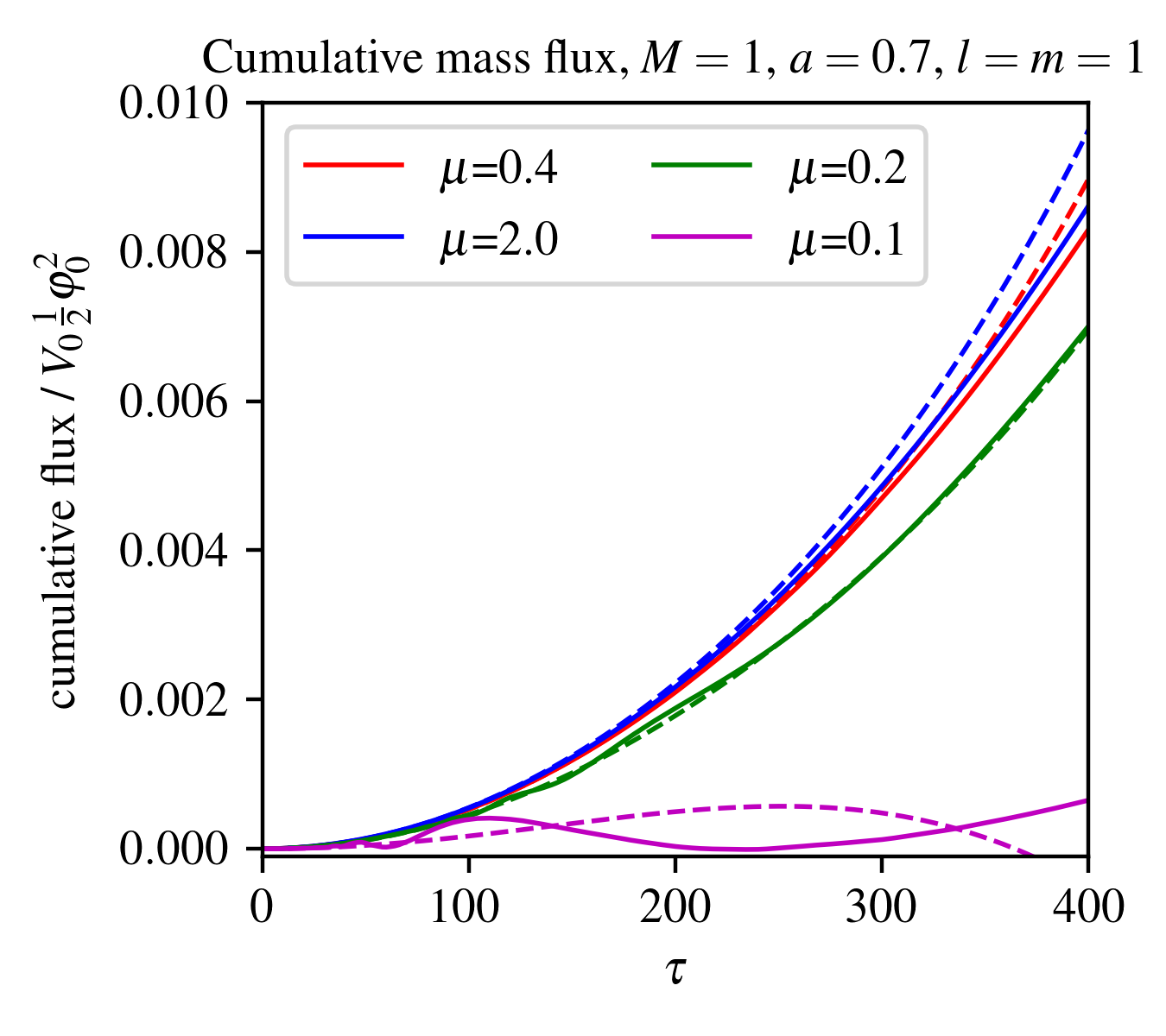}
    \caption{Cumulative flux into a sphere of $R=300M$ for $M=1,\chi=0.7,\alpha=0,l=m=1$ and different $\mu$, normalised by $\tfrac{1}{2}V_0\varphi^2_0 = E_0/\mu^2$. This again shows good agreement with the perturbative result. We see that the very low mass cases show an oscillating behaviour in the flux as a result of the stationary wave profiles in this regime.}
    \label{fig:compare_mu1}
\end{figure}

Fig.~\ref{fig:compare_mu1} shows the effect of changing the scalar field mass $\mu$ for fixed $\chi=0.7$, and $l=m=1$. We see that for $M\mu=0.4, 2.0$ the simulation flux again deviates from the perturbative expression at roughly $\tau \sim 300$, however the small mass case $M\mu=0.1$ shows deviation at much smaller $\tau$. If we examine the perturbative analytic series \eqref{eq:mass_flux_perturb} expressed in terms of $\tau$ we see that $\mu$ enters chiefly as $\Tilde{\Lambda}=l(l+1)/(M\mu)^2 \sim (l/\mu)^2$ (it also appears as $\mu^{-2}$ at order $\mathcal{O}(\Tilde{r}^{-5})$ and above). Decreasing $M\mu$ with fixed $l,m$ corresponds to increasing $\Tilde{\Lambda}$ and thus the angular momentum per unit mass, which again leads to a decreased growth rate. In terms of the perturbative expansion a larger $\Tilde{\Lambda}$ boosts the effect of higher order terms and causes the perturbative solution to break down at smaller $\tau$. Physically, the oscillatory behaviour of the flux is a result of the stationary wave profiles that develop in this mass regime.

\begin{figure}[h!]
    \centering
    \includegraphics{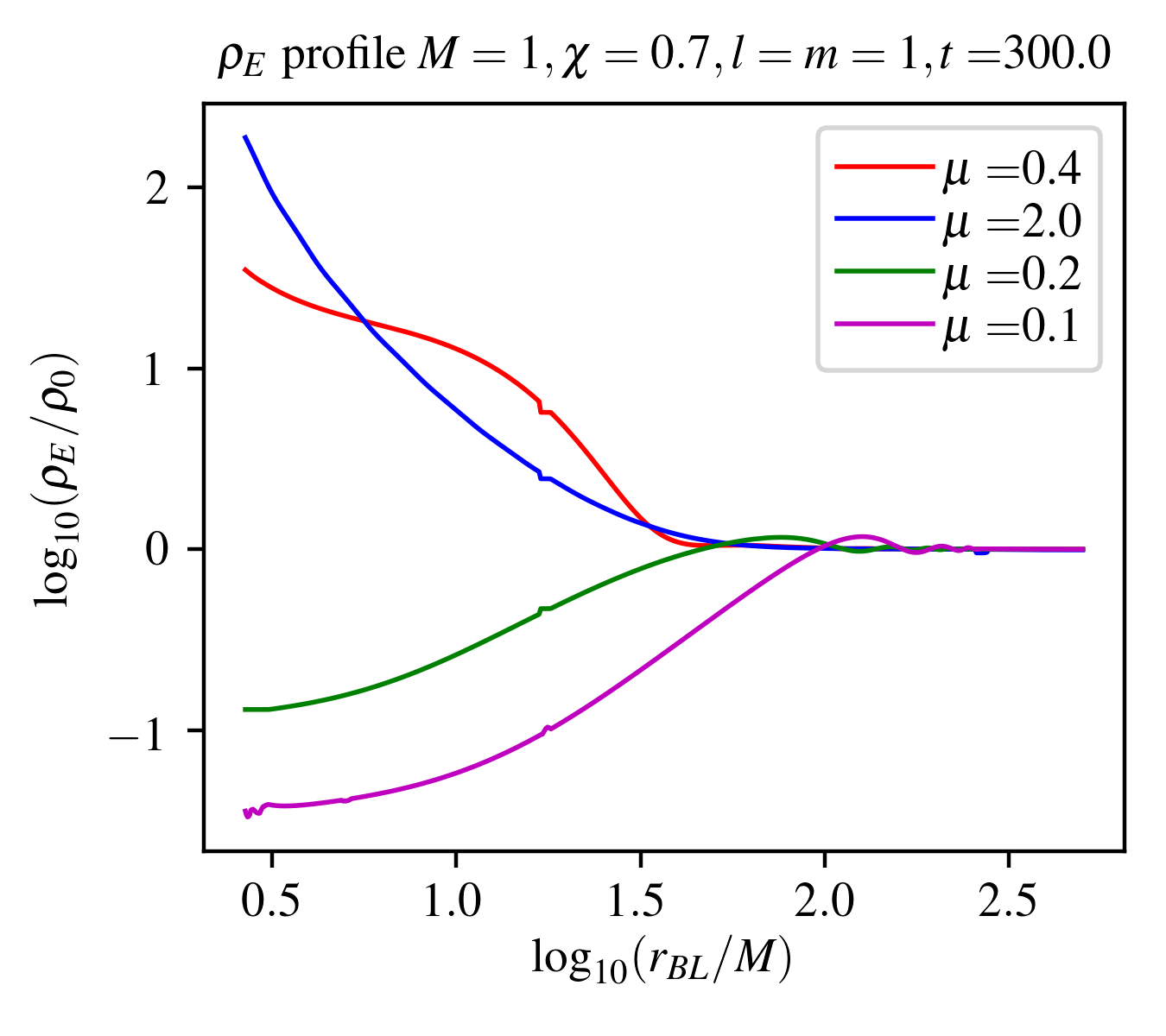}
    \caption{Radial profile of the $0,0$ harmonic mode of the energy density $\rho_E$, i.e. the energy density is averaged over a sphere at each radius. The initial scalar field parameters are $M=1,\chi=0.7,l=m=1$ and different $\mu$. We find the maximum density occurs for the highest mass cases where we have accretion onto the BH. The results are shown at time $t=300M$ which corresponds to $\tau=120$ for $\mu=0.4$.}  \label{fig:rho_profile_compare_mu_vs_t}
\end{figure}

Fig. \ref{fig:rho_profile_compare_mu_vs_t} shows that the maximum density occurs for the highest mass cases where we have accretion onto the BH. For lower $\mu$ the cloud is concentrated further from the BH and the energy is thus more diluted. As in the superradiant case we see that the scalar mass $M\mu\sim0.4$ is a critical value where we still support a cloud outside the horizon, rather than having an accretion flow all the way to the horizon.

\subsubsection{Effect of alignment angle $\alpha$}

\begin{figure}
    \centering
    \includegraphics{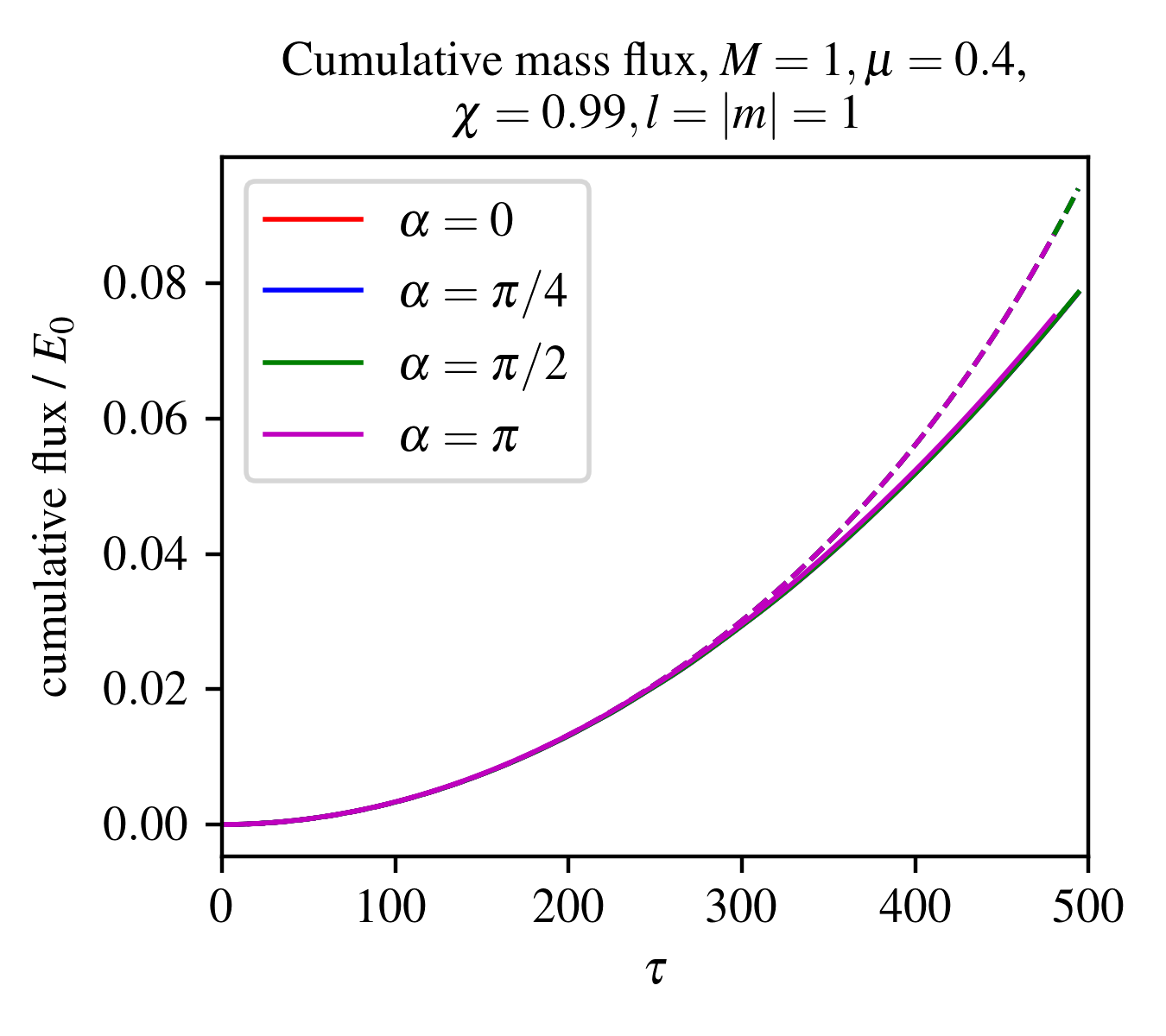}
    \caption{Cumulative accretion flux of mass into a sphere of $R=300M$ around a Kerr black hole for $\mu=0.4,\chi=0.99,l=|m|=1$ and different alignment angles $\alpha$ (the analytic result is given by equation \eqref{eq:alignment_angle}). At this large radius, we see little difference in the accretion rate towards the BH for different $\alpha$.}
    \label{fig:compare_Al}
\end{figure}

\begin{figure}
    \centering
    \includegraphics{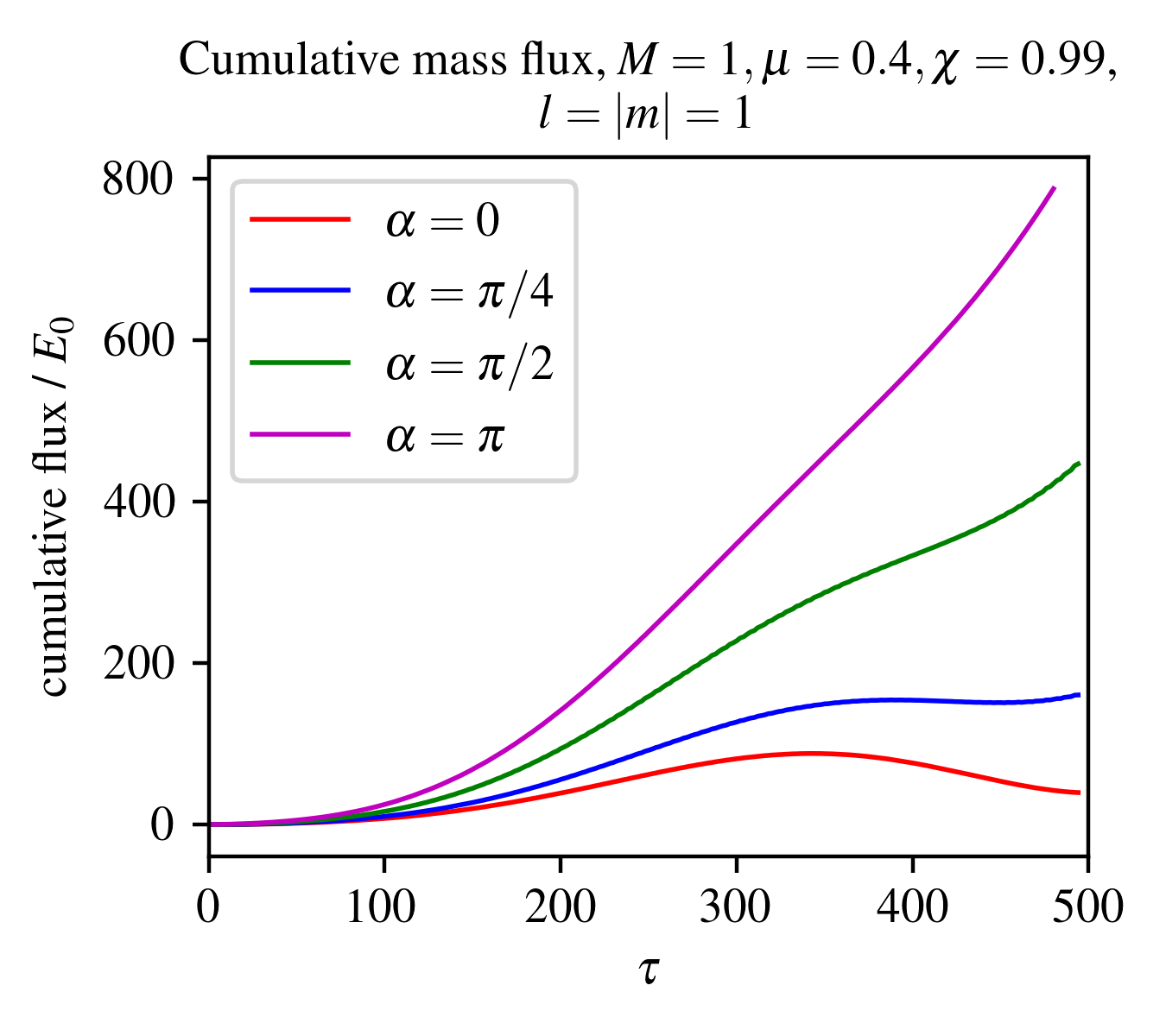}
    \caption{Cumulative accretion flux of mass into a sphere of $R=10M$ around a Kerr black hole for $\mu=0.4,\chi=0.99,l=|m|=1$ and different $\alpha$. At this radius we clearly see the effect of the misalignment in increasing the flux towards the horizon. Note we cannot apply the perturbative analytic result in equation \eqref{eq:alignment_angle} at this small radius.}
    \label{fig:compare_Al_R10}
\end{figure}

\begin{figure}
    \centering
    \includegraphics{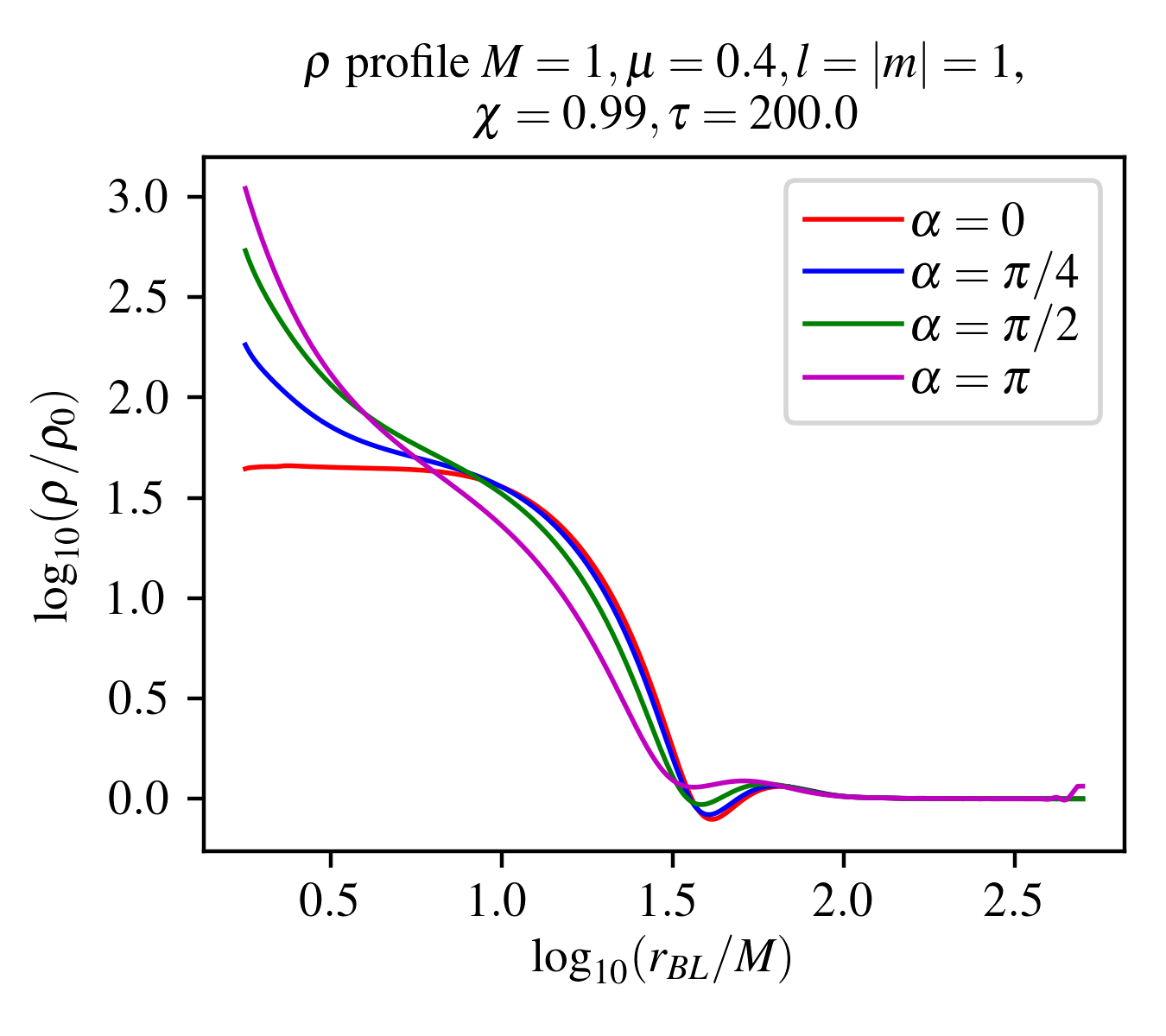}
    \caption{Radial profile of the $0,0$ harmonic mode of the energy density $\rho_E$, i.e. energy density is extracted by averaging over a sphere at each radius. The parameters are $M\mu=0.4,\chi=0.7,l=|m|=1$ and different alignment angle $\alpha$. We see that the density near the BH is enhanced for misaligned spins, indicative of the accretion flow.}  \label{fig:rho_profile_compare_Al}
\end{figure}

\begin{figure*}[]
    \centering
    \subfigure[$\; \alpha=0$ \label{fig:alpha_0}]{
            \includegraphics[width=0.35\textwidth]{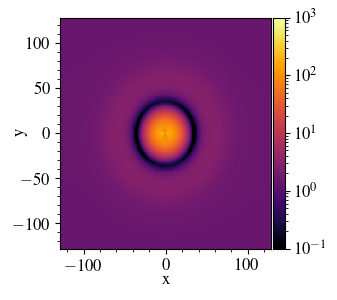}}
    \subfigure[$\; \alpha=\pi/4$ \label{fig:alpha_0.25pi}]{
            \includegraphics[width=0.35\textwidth]{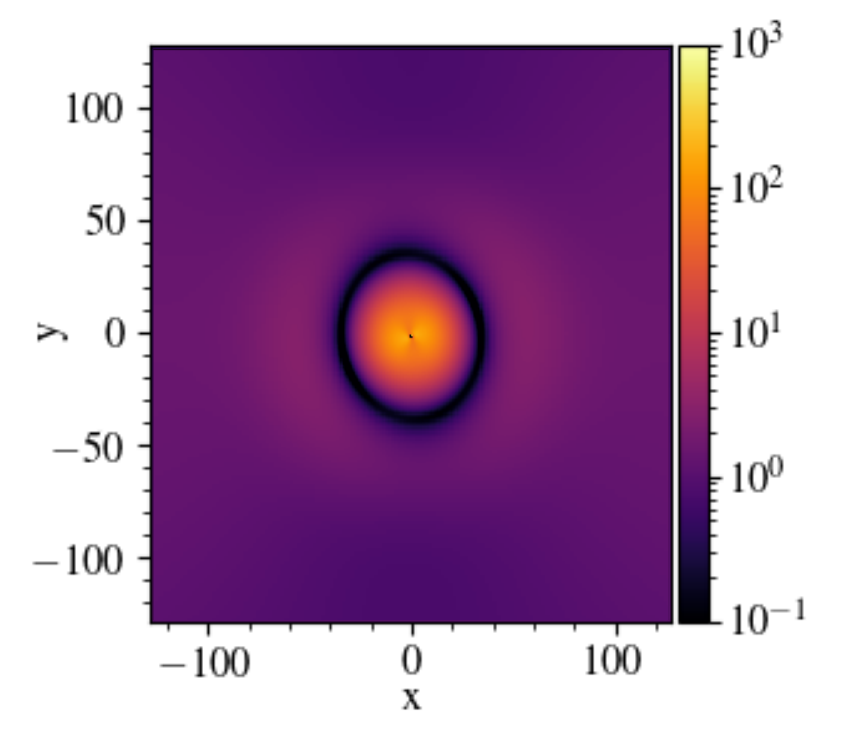}}
    \subfigure[$\;  \alpha=\pi/2$ \label{fig:alpha_0.5pi}]{
            \includegraphics[width=0.35\textwidth]{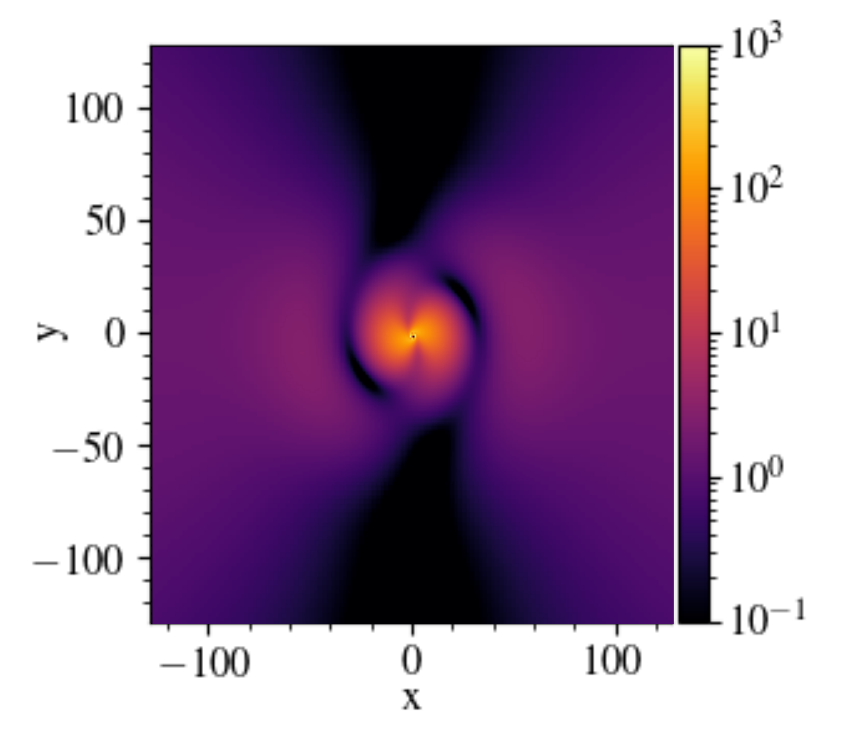}}
    \subfigure[$\; \alpha=\pi$\label{fig:alpha_pi}]{
            \includegraphics[width=0.35\textwidth]{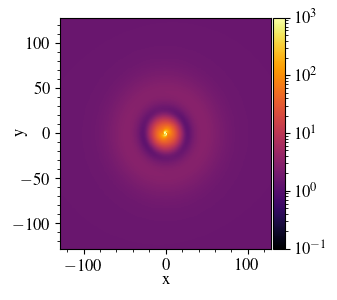}}
   \caption{Pseudo-colour plots of $\log_{10}(\rho_E/\rho_0)$ of the scalar field in the $z=0$ plane for $\chi=0.99,\mu=0.4,\tau=320$, an initial $l=m=1$ angular profile and different alignment angle $\alpha$ between the BH spin and scalar angular momentum.}
    \label{fig:alpha_rho_plots}
\end{figure*}

Here we vary the alignment angle $\alpha$\footnote{Not to be confused with lapse $\alpha$ or the $\alpha$ parameter in the confluent Heun function.} between the BH and cloud spin, fixing the rest of the parameters. Fig.~\ref{fig:alpha_rho_plots} shows 2D profiles of the energy density in the $z=0$ plane for different $\alpha$, with fixed $\chi=0.99$, $\mu=0.4$, $M=1$ and initial $l=m=1$ angular dependence.
We can see that changing $\alpha$ does produce significant differences in the profiles around the BH.

However Fig.~\ref{fig:compare_Al} shows that changing  $\alpha$ has only a very small effect on the total flux at a larger radius. In this figure, the solid lines describe the numerical results on the time evolution of the total flux, and the dashed lines describe the analytical estimates. As expected, the analytical solution describes the numerics well only during early times. 

In particular, for $l=|m|=1$ the perturbative solution gives 
\begin{equation}\label{eq:alignment_angle}
\begin{split}
    \delta M_{\textup{cloud}} - \delta M_{\textup{cloud}}\Big\rvert_{\alpha=0} &= 4\pi\Phi^2_0 \frac{a^2\tau^2\sin^2(\tfrac{1}{2}\alpha)}{\Tilde{r}^2}\bigg\{3\Tilde{m}\\ &-\frac{3}{5}\cos^2(\tfrac{1}{2}\alpha)\\
    &-\frac{10\Tilde{m}-2\cos^2(\tfrac{1}{2}\alpha)}{\Tilde{r}}+\mathcal{O}(\Tilde{r}^{-2})\bigg\}
\end{split}
\end{equation}
where $\Tilde{m}=\chi/(M\mu)$ and again neglecting the oscillating terms. Here we see the change in $\delta M_{\textup{cloud}}$ due to $\alpha$ in \eqref{eq:alignment_angle} is proportional to $a^2$, and we can thus interpret it as the spinning black hole exerting ``friction" on the scalar field and removing its angular momentum via frame dragging, making it easier for the scalar field particles to fall inwards. As the perturbative expression is proportional to $1/\Tilde{r}^2$ we would expect this effect to increase at smaller $r$. Fig. \ref{fig:compare_Al_R10} shows the mass flux into a sphere at a smaller radius of $R=10M$, and the increase in growth rate on increasing $\alpha$ is now clearly visible. 

We also saw in Fig.~\ref{fig:Kerr_eff_pot} how changing from $m=1$ (aligned i.e.~$\alpha=0$) to $m=-1$ (anti-aligned i.e.~$\alpha=\pi$) causes the potential barrier to vanish close the the BH, whilst further out the potentials were similar. Examination of Figs.~\ref{fig:alpha_rho_plots} and \ref{fig:rho_profile_compare_Al} shows that for the $\alpha=\pi$ case the energy density in the $z=0$ plane is more concentrated close to the BH vs the $\alpha=0$ case, which is indicative of accretion onto the BH.

\clearpage

\subsubsection{Comparison of angular momentum and mass growth rates}

\begin{figure}
    \centering
    \includegraphics{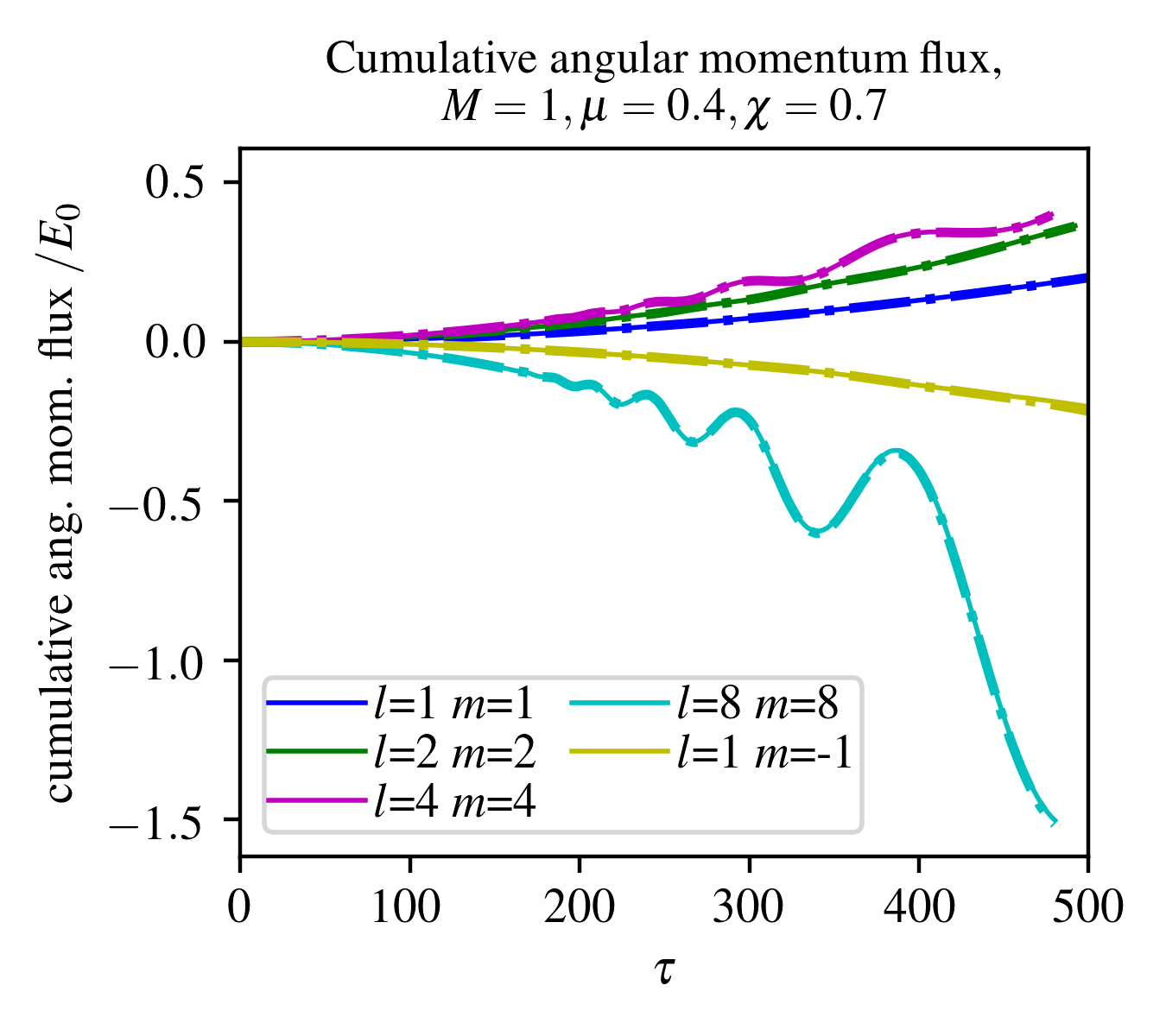}
    \caption{Cumulative angular momentum flux into a sphere of $R=300M$ (solid line) and the mass flux multiplied by $m/\mu$ (dot dashed line) for $\chi=0.7, \mu=0.4$ and different $l,m$. We see that the solid and dot dashed lines agree closely, indicating that the angular momentum per unit mass is approximately constant at $m/\mu$.}
    \label{fig:compare_ang_mom_lm}
\end{figure}

We can perform the same measurements for the conserved total angular momentum and the angular momentum flux. Fig.~\ref{fig:compare_ang_mom_lm} shows the angular momentum flux for different $l,m$ and the mass flux from Fig.~\ref{fig:compare_lm}  multiplied by $m/\mu$. They agree closely, which confirms what we found from the perturbative solutions in section \ref{sec-analytic} that at $r \gg M$ the angular momentum per unit mass is approximately constant at $m/\mu$. This is what we would expect if we model the scalar field as a collection of non-interacting classical particles which each individually conserve angular momentum. 

\begin{figure}
    \centering
    \includegraphics{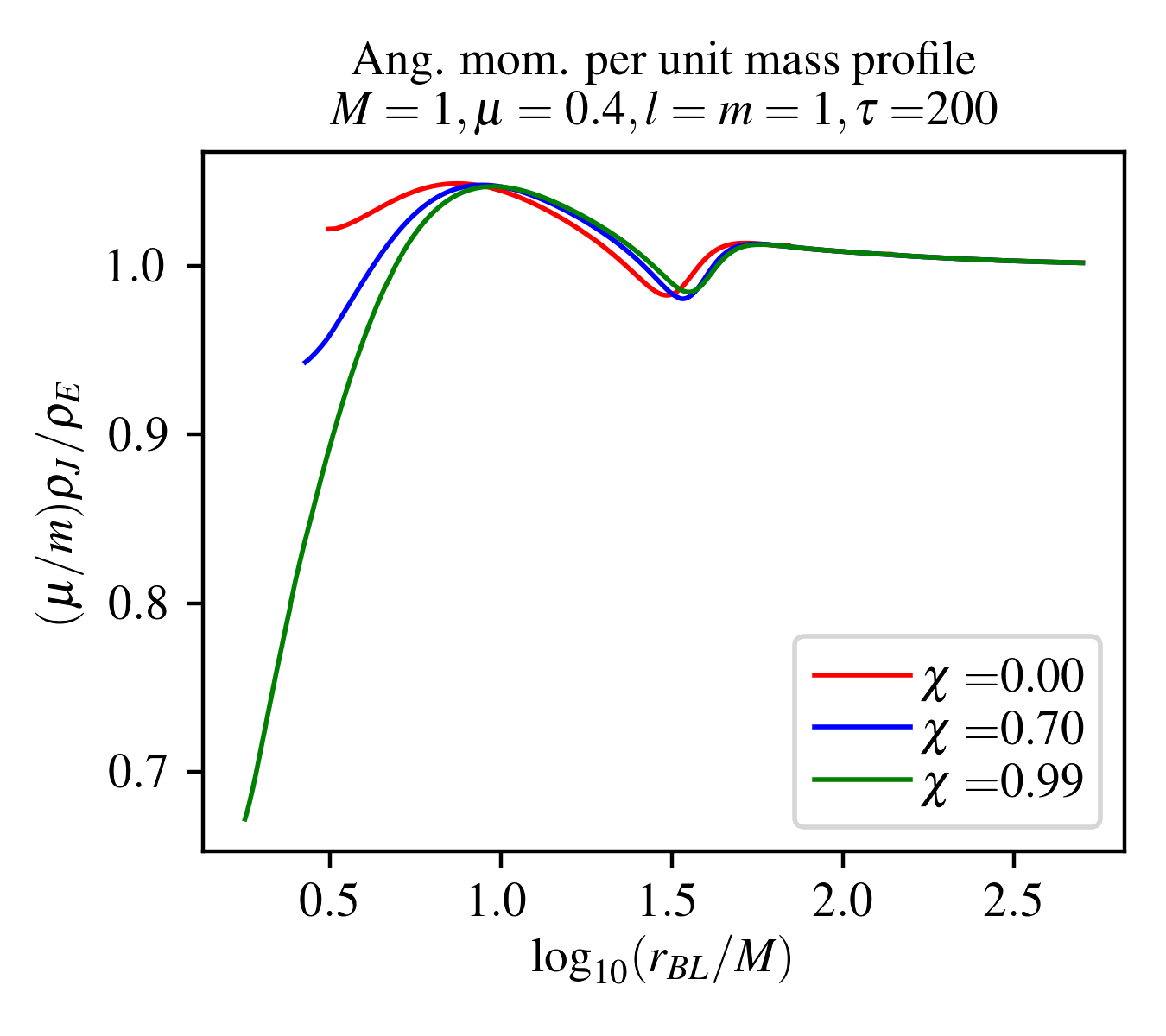}
    \caption{Angular momentum per unit mass, $\rho_J/\rho_E$, divided by $m/\mu$ vs the BL radius at $\tau=200$ for different BH spin. We see that the frame dragging effect serves to increase the angular momentum close to the more highly spinning BHs.}
    \label{fig:ang_mom_profile}
\end{figure}

Fig.~\ref{fig:ang_mom_profile} shows the ratio of the angular momentum density to mass density $\rho_J/\rho_E$ (each density averaged over the sphere) vs radius divided by $m/\mu$. Again we see that at large $r$ this value approaches 1, indicating $\rho_J/\rho_E \approx m/\mu$, however at smaller $r$ close to the horizon we see a distortion, which increases with increasing BH spin. We can interpret this as the BH frame dragging effect. 

\section{Gravitational waves}
\label{sec-gws}

For minimally coupled scalar hair without a direct coupling to Standard Model matter, the only way of detecting its presence is through its gravitational effects. Gravitational wave observations may then act as a probe of such scalars, for example, via the impact of a high density cloud on compact binary merger signals, or from the gravitational decay of the cloud itself. 

In the latter case, it is known that a rotating non-axisymmetric mass distribution can give rise to a quasi-monochromatic gravitational wave signal at twice the oscillation frequency \cite{Arvanitaki_2015}. This signal can also be considered as arising from the annihilation of two scalar bosons to produce gravitons in the background of the black hole. Gravitational waves can also be produced from transitions between energy states of the scalar field \cite{Arvanitaki_2015}. Searches for such signals in relation to superradiant clouds have been proposed and attempted using advanced LIGO and Virgo \cite{Palomba_2019, Sun_2020, Isi_2019, Yoshino_2014, Ken_2020, Ken_2020constraints, Ken_2020_3} and explored for future detectors such as LISA \cite{Ghosh_2017, Berti_2019, Arvanitaki_2011, Barausse:2020rsu, Baibhav:2019rsa}. Their absence provides observational constraints on the existence of light massive real scalar fields in particular mass ranges.

Real scalar clouds formed from gravitational accretion that we study here may give rise to a similar effect due to the axisymmetric mass distribution of the clouds formed - as illustrated in Fig.~\ref{fig:rho_close_up}. To the best of our knowledge, it has not previously been suggested that the simple accretion of scalar matter around BHs could give rise to such signals. In this section we therefore make a rough calculation of the size of the signal.

We will again consider marginally bound states with $\omega=\mu$. We first consider the scalar-scalar annihlation signal where the scalar field is dominated by a single $l=m$ mode. The size of the scalar cloud is determined by the dark matter environment. Following Hui et al. (2019) \cite{Hui_2019} we take the size of the cloud $R_c$ to correspond to the radius of influence of the black hole 
\begin{equation}
    R_c \sim r_i \sim M/v^2 \sim 10^7 M \left(\frac{v_0}{v_{\phi}}\right)^2,
    \label{eq:typical_R_c}
\end{equation}
where $v_{\phi}$ is the typical axial velocity of the scalar field particles at $R_c$ and $v_0 = 100 \textup{km s}^{-1}$ is taken to be a typical velocity scale for dark matter \cite{Hui_2019}. 

We make the simplifying assumption of a single mode and assume the mode number is set by the average axial velocity. The angular momentum per unit mass for a $l=m$ mode is approximately $m/\mu$ so
\begin{equation}
\begin{split}
    m/\mu &\sim v_{\phi} R_c, \\
    m &\sim v_{\phi} \mu R_c.
    \label{eq:typical_m}
\end{split}
\end{equation}
Let us consider a value of $\mu$ such that $m \sim 1$. Let $\alpha_g = M\mu$ be the dimensionless ratio of BH radius to scalar wavelength. Then
\begin{equation}
    \alpha_g \sim m v_{\phi} \sim 10^{-3} (v_{\phi}/v_0),
    \label{eqn:alphag}
\end{equation}
so $v_{\phi} \sim v_0$ gives $\alpha_g \sim 10^{-3}$ and $R_c/M \sim 10^7$. The simplest way to estimate the gravitational wave emission is the quadrupole formula
\begin{equation}
    L_{GW} \approx \frac{1}{5}\left\langle \pdv[3]{Q^{TT}_{ij}}{t}\pdv[3]{Q^{TT}_{ij}}{t}\right\rangle.
\end{equation}
where $Q^{TT}_{ij}$ is the quadrupole moment in the transverse traceless gauge. However this relies on the approximation that the size of the source is much smaller than the gravitational wave wavelength \cite{Kokkotas_2008}. The ratio of cloud size to wavelength is 
\begin{equation}
    R_c/\lambda_c \sim R_c\mu \sim 10^3 (v_{\phi}/v_0)^{-1} \gg 1,
\end{equation}
so for $v_{\phi} \sim v_0$ the quadrupole approximation is not appropriate. The authors in \cite{Brito_2015} use the Teukolsky formalism to go beyond the quadrupole formula and derive that for small $\alpha_g$ and the bound $n=0$, $l=m=1$ mode, the GW luminosity is approximately 
\begin{equation}
    L_{GW} \sim M^2_s \alpha^{12}_g \mu^2 \label{eq:Brito_formula}
\end{equation}
where $M_S$ is the mass of the scalar cloud. We can estimate that 
\begin{equation}
    M_S \sim \rho_{R_c} R^3_c
\end{equation}
where $\rho_{R_c}$ is the scalar field energy density at the radius of influence $R_c$. Then
\begin{equation}
    h_{GW} = \frac{2\sqrt{L_{GW}}}{d(2\mu)} \sim \frac{\rho_{R_c} R^3_c \alpha^6_g}{d} \sim \frac{\rho_{R_c}M^3}{d},
\end{equation}
where $d$ is the distance from the source to the detector. We can reintroduce the constants to obtain
\begin{align}
    h_{GW} &\sim \left(\frac{G}{c^2}\right)^4 \frac{\rho_{R_c} M^3}{d}, \\
    &\sim 10^{-53} \left(\frac{\rho_{R_c}}{M_{\odot} \pc^{-3}}\right) \frac{(M/M_{\odot})^3}{(d / \pc)}.
\end{align}
As a concrete example consider the supermassive black hole at the centre of the Milky Way. Various studies have estimated the dark matter density profile of the Milky Way's halo from the galaxy rotation curve (see \cite{sofue2020rotation} for a review). Estimating the DM density at the centre of the galaxy is difficult as the mass is dominated by baryonic stars, and estimates are highly model dependent. Light scalar dark matter predicts a solitonic ``core" of almost uniform density near the centre \cite{Li_2020}. Nesti and Salucci (2013) \cite{Nesti_2013} estimate a core density of $\rho_c \sim 0.04 M_{\odot} pc^{-3}$, and a core radius of $\sim 10$kpc which appear to be typical values for cored models. Using $M = 4\times 10^{6} M_{\odot}$ for the mass of the supermassive black hole \cite{Nesti_2013}, $\rho_{R_c}=\rho_c$, and $d = 8$kpc for the distance to the centre of the Milky Way we obtain
\begin{equation}
    h_{GW} \sim 10^{-39}.
    \label{eq:h_estimate}
\end{equation}
The corresponding scalar mass and GW frequency are 
\begin{align}
    \hbar\mu &\sim \hbar \left(\frac{m v_0 c^2}{G M}\right)\left(\frac{v_{\phi}}{v_{\phi}}\right)(v_{\phi}/v_0) \\
    &\sim 10^{-20}\kappa^{-1}(v_{\phi}/v_0) ~ \eV, \\
    f &\approx c\mu/\pi \sim 10^{-5}(v_{\phi}/v_0) ~ \textup{Hz},
\end{align}
and the size of the cloud is 
\begin{equation}
    R_c \sim (v_{\phi}/v_0)^{-2} ~ \pc.
\end{equation}
If we take $v_{\phi} \sim v_0$ then the frequency is outside the range of LIGO \cite{LIGO} and on the edge of the sensitivity range of LISA \cite{LISA,LISA_sens}. The signal of $h_{GW} \sim 10^{-39}$ would also be far below the threshold of the latter even at peak sensitivity \cite{LISA_sens}. Note that this estimate would be further reduced by the fact that the DM is likely to be in a superposition of modes rather than a single coherent one. Moreover, the formula \eqref{eq:Brito_formula} was derived for the $n=0$ mode, which goes to zero at $r=\infty$, while we are examining the marginally bound or $n=\infty$ mode, which is finite at $r=\infty$, so we would expect this to be suppressed due to the smaller value of $\tilde r = 2 r M \mu^2
/(1 + n + l)$ and the more spread out profile. We leave a recalculation of the \cite{Brito_2015} result for the appropriate $n$ to future work.

Equation \eqref{eq:typical_m} suggests increasing $\mu$ would increase the typical $m$ in which case other modes may dominate. Arvanitaki et al. (2015) \cite{Arvanitaki_2015} found that the GW emission rate from single mode scalar-scalar annihilation in bound gravitational atom states (with $\omega < \mu$) goes as 
\begin{equation}
    \Gamma_{GW} \propto \left(\frac{v_{\phi}}{c/2}\right)^{4l}.
\end{equation}
Hence as $c \gg v_{\phi} \sim 100 \textup{km s}^{-1}$ we expect the emission to decrease for larger $l$ (assuming the same occupation number of the respective modes). Therefore for single mode annihilation we may consider the $l=m=1$ case to be the most optimistic scenario.  

We can also consider radiation arising from transitions between the marginally bound state at $\omega=\mu$ and lower energy states with $\omega < \mu$. These transitions will produce radiation at lower frequencies - in particular, if the average axial velocity of the scalar particles at $r=R_c$ is $v_{\phi}$ then we expect radiation at frequency $
\omega \sim v_{\phi}/R_c \sim \mu v_{\phi}^2 /m$. Bound states for small $\alpha_g$ have energies 
\begin{equation}
    \omega_{nl} \approx \mu - \frac{\mu}{2}\left(\frac{\alpha_g}{n+l+1}\right)^2,
\end{equation}
where $n$ is a non-negative integer and related to the number of nodes in the radial direction. Consider for example a transition from a marginally bound $n=\infty,l=m=2$ state to the ground state $n=l=m=0$. This would produce radiation of frequency 
\begin{equation}
    \omega_{GW} \sim \frac{\mu}{2}m^2 v^2_{\phi} = 2 \mu v^2_{\phi}.
\end{equation}
Due to the lower frequency of the GWs the ratio of cloud size to wavelength is now
\begin{equation}
    R_c/\lambda \sim 10^{-3} (v_{\phi}/v_0) \ll 1,
\end{equation}
so we can use the quadrupole formula to estimate the gravitational radiation, 
\begin{align}
    L_{GW} \sim& ~ \omega^6_{GW} \left(\int^{R_c}_{r_+} \rho r^4 \dd r\right)^2, \\
    h_{GW} \sim& ~ \frac{2\sqrt{L_{GW}}}{d\omega_{GW}} \sim \omega^2_{GW} \int^{R_c}_{r_+} \rho r^4 \dd r \\
    \sim& \left(\mu v^2_{\phi}\right)^2 \frac{\rho_{R_c} R^5_c}{d},  \\
    \sim& 10^{-40} \left(\frac{\rho_{R_c}}{M_{\odot} \pc^{-3}}\right) \frac{ (M/M_{\odot})^3}{(d / \pc)(v_{\phi}/v_0)^4}. 
\end{align}
Using the same values for the Milky Way SMBH as above this gives an amplitude of
\begin{equation}
    h_{GW} \sim 10^{-26} (v_{\phi}/v_0)^{-4}.
\end{equation}
with the corresponding scalar mass and GW frequency
\begin{align}
    \hbar\mu \sim& ~ \hbar \left(\frac{m c^2}{G M}\right)\left(\frac{v_{\phi}}{v_0}\right) \sim
    10^{-20}(v_{\phi}/v_0)\eV, \\
    \quad f \sim& ~ 10^{-12}(v_{\phi}/v_0)^3\textup{Hz}.
\end{align}
This frequency would be well outside the range of LIGO and LISA but may be in reach of Pulsar Timing Arrays \cite{PTA_review}. Note again that this estimate is based on the strong assumption that the DM is all in the single mode, whereas in reality only some fraction of the total will be. 

The GW emission estimates presented here are only rough guides for a single nearby source. Accurate estimates for the emission of GWs through either annihilation or level transitions would require a more detailed calculation similar to those in \cite{Brito_2015} and \cite{Khmelnitsky:2013lxt}. In addition as the mode profile is heavily determined by the scalar environment, ideally one would like to obtain more precise information about the central distribution of the DM in order to construct a more reliable estimate of the complete signal.

If dark matter is a scalar with a single mass and a relatively consistent velocity profile, both signals should be largely monochromatic. One could therefore expect a superposition of signals with a similar frequency to arise from multiple black holes in the observable volume. This could potentially lead to an enhancement in the total GW signal. Estimates of such a stochastic quasi-monochromatic GW background have been obtained in the context of searches for superradiant clouds \cite{Arvanitaki_2015,Ghosh_2017,Palomba_2019}.

\section{Discussion}
\label{sec-discussion}
Most of the literature concerning the growth of scalar hair around black holes has focused on the superradiance mechanism. This work instead explores simple gravitational accretion driven growth of scalar fields around black holes. Specifically we have developed analytic tools to characterise the growth as a function of a range of parameters of the BH and scalar, and performed simulations of the field evolution on a fixed Kerr background to validate our results. If dark matter is composed of light bosons this should represent a common environment for astrophysical BHs.

We observed that when one includes a spin for the BH, as would be expected in a realistic astrophysical case, the accretion rates and density profiles remain almost unchanged, with only minor distortion near the horizon. Hence the behaviour of accretion onto Kerr BHs remains fundamentally very similar to that onto Schwarzschild BHs studied in \cite{Clough_2019}, with the profile ``spiking'' around the horizon. Over time the field profiles come to resemble the analytic stationary solutions described by \cite{Hui_2019}. 

However, adding angular momentum to the scalar can either suppress or enhance (depending on the misalignement) its accretion onto the BH and, in the case of aligned spins, concentrates the clouds further out from the horizon. This is interesting for two reasons:
\begin{enumerate}
    \item The specific profile which forms around the BH is important because it would directly affect potential probes of the cloud structure - e.g. dephasing in EMRIs by LISA.
    \item The flux onto the BH determines how fast the BH may be spun up or how fast its mass will grow. This may have implications for the superradiant growth that would be expected to accompany the accretive growth in several regions of the parameter space studied. 
\end{enumerate}
Regarding the second point, it would be interesting to consider whether this competition between spin up and spin down could stall or enhance the superradiant build up in some cases, in a similar way to the study in \cite{Ficarra:2018rfu}.

We have explained the cloud behaviour by reference to a quasi-effective radial potential and the orbits of equivalent particles, and developed a perturbative analytic solution for the changing field profile at large radius. This proved very effective in describing the mass and angular momentum accretion flux in the appropriate time range, and helped us understand the behaviour in the full numerical simulations.

We have not considered couplings to Standard Model matter in this paper, or self interactions in the scalar field, but our work could be used to inform estimates of potential signals from such effects \cite{Yuan:2020xui}. In the absence of such couplings, the key observational signature of the scalar field cloud would be gravitational waves. We are currently studying the impact of these clouds on a binary merger waveform, in simulations which include the backreaction effect on the metric, and so we leave comments on that case to future work. However, an interesting feature of the scalar clouds with angular momentum that we study here is that they should generate monochromatic GW signatures similar to those proposed in the case of superradiant clouds.
We have provided a rough estimate for the size of signal expected from a cloud around the supermassive BH at the centre of the Milky Way. This indicates that a real scalar DM cloud around a single BH would not produce a GW signal detectable by any planned GW observatories. However, the strong dependence on the asymptotic conditions and the potential for a superposition of multiple signals motivates a more detailed specification of the DM environment surrounding black holes.

\section*{Acknowledgements}
\vspace{-0.2in}
\noindent JB and KC thank the GRChombo collaboration (www.grchombo.org) for their support and code development work. We thank V Cardoso, R Croft, S Ghosh, M Radia and H Witek for helpful conversations. JB acknowledges funding from a UK Science and Technology Facilities Council (STFC) studentship.
PF and KC acknowledge funding from the European Research Council (ERC) under the European Unions Horizon 2020 research and innovation programme (grant agreement No 693024). During the development of this work, ML was funded by the Kavli Institute for  Cosmological  Physics  at  the  University  of  Chicago through an endowment from the Kavli Foundation and its founder Fred Kavli. ML was also funded by the Innovative Cosmology Theory fellowship at the University of Columbia. LH was supported by a Simons Fellowship and the Department of Energy DE-SC0011941.
The simulations presented in this paper used the Glamdring cluster, Astrophysics, Oxford, and DiRAC (Distributed Research utilising Advanced Computing) resources under the project ACSP218. This work was performed using the Cambridge Service for Data Driven Discovery (CSD3), part of which is operated by the University of Cambridge Research Computing on behalf of the STFC DiRAC HPC Facility (www.dirac.ac.uk). The DiRAC component of CSD3 was funded by BEIS capital funding via STFC capital grants ST/P002307/1 and ST/R002452/1 and STFC operations grant ST/R00689X/1. DiRAC is part of the National e-Infrastructure.
The authors also acknowledge the computer resources at SuperMUCNG and the technical support provided by the Leibniz Supercomputing Center via PRACE (Partnership for Advanced Computing in Europe) Grant No. 2018194669.

\appendix

\section{Code tests and validation}\label{App:Flux}

In this Appendix we discuss how we validate our code results, and provide some additional details on the formulation of problems in fixed metric backgrounds. 

\subsection{Code validation and coordinate choice}

As discussed in the main text, we evolve the field on a fixed background Kerr metric in Quasi Isotropic Kerr (QIK) coordinates.

The metric is validated by checking that the numerically calculated Hamiltonian and Momentum constraints converge to zero with increasing resolution, as do the time derivatives of the metric components, i.e. $\partial_t \gamma_{ij} = \partial_t K_{ij} = 0$ (calculated using the ADM expressions). This ensures that the metric which is implemented is indeed stationary in the chosen gauge, consistent with it being fixed over the field evolution. 
The outer horizon at $R = r_+$ is spherical and
therefore it retains a finite limit even in the limit $\chi \rightarrow 1$. However, as noted above, the use of QIK coordinates necessitates the use of an analytic continuation of the lapse in which its value becomes negative within the horizon. To see why this is useful in comparison to the positive continuation, note that the solution within the horizon describes a mirror universe, rather than a BH interior, so this choice corresponds to running time ``backwards'' in this region. As a consequence any matter will fall towards the grid centre, i.e. towards asymptotic spatial infinity in the mirror universe. One does not then in principle need to excise within the horizon, but in practise we do stil excise some part of the interior to prevent any spikes developing.

The advantage of these coordinates is their simple relation to the BL coordinates in which we perform our perturbative analysis. The downside is that since they correspond to the asymptotic observers, the lapse goes to zero and time freezes around the horizon. Thus ingoing waves tend to ``bunch up'' there. Given sufficient resolution outside the outer horizon, the ingoing nature of the metric prevents the errors this introduces from propagating into the region far from the BH, and unresolved waves are effectively damped away by grid precision close to the horizon. Provided we are not interested in extracting quantities very close to the horizon, these coordinates work in practise. 

An alternative set of coordinates are Kerr-Schild coordinates \cite{KerrSchild}. This form has the advantage of being horizon penetrating, in other words without a coordinate singularity at the horizon. The metric can be split into 
\begin{equation}
g_{\mu \nu} = \eta_{\mu \nu} + 2H(x^{\mu})l_{\mu}l_{\nu},
\end{equation}
where $\eta_{\mu \nu}$ is the Minkowski metric, $H = Mr/\Sigma$, and $l^{\mu}$ is an ingoing null vector given by
\begin{equation}
l^{\mu} = \left(-1, 1, 0, -\frac{2a}{a^2 + r^2}\right),
\end{equation}
written out in coordinates $\{t, r, \theta, \phi\}$. These relate to the Boyer-Lindquist coordinates via
\begin{align}
    t_{KS} &= t_{BL} + \frac{2M}{r_+ - r_-}\left(r_+\ln \left\vert \frac{r}{r_+} - 1\right\vert - r_- \ln \left\vert \frac{r}{r_-} - 1\right\vert\right), 
    \label{eq:t_KS_BL}\\
    \phi_{KS} &= \phi_{BL} + \frac{a}{r_+ - r_-}\ln\left\vert \frac{r - r_+}{r - r_-}\right\vert + 2\tan^{-1}\left(\frac{a}{r}\right).
    \label{eq:phi_KSBL}
\end{align}
Taking $a=0$ one can show these coordinates reduce to ingoing Eddington-Finkelstein (EF) coordinates rather than Schwarzchild, with ingoing EF null coordinate $v = r + t_{KS}$. This makes it more difficult to interpret results expressed in Kerr-Schild coordinates from the perspective of a distant observer as $t_{KS}$ is no longer their measured time at finite $r$. 
Taking $a = 0$ and setting $u = t+r$ one can show that the radial coordinate $r$ corresponds to that of ingoing Eddington-Finkelstein rather than Schwarzschild coordinates. This makes it more difficult to interpret results expressed in Kerr-Schild coordinates from the perspective of a distant observer as $t_{KS}$ is no longer their measured time at finite $r$.
For simplicity of presentation we have presented results in Quasi Isotropic Kerr coordinates only, but we have checked that we obtain physically consistent results in both gauges, and that our diagnostics can be reconciled in both cases, see for example Fig. \ref{fig:KS_mass_flux_agreement}.

\subsection{Conserved Fluxes}

In numerical relativity it is conventional to decompose the stress energy tensor into purely spatial quantities as
\begin{equation}
	T_{\mu\nu} = \rho n_\mu n_\nu + S_\mu n_\nu + n_\mu S_\nu + S_{\mu\nu}
\end{equation}
Then $\rho$ and $S^i$ are the energy and momentum densities measured by the Eulerian observers, that is, observers moving normal to the spatial hypersurfaces in the ADM decomposition. These are not the same as the time-like observers for which the conserved quantities are defined. We can obtain the expressions for the conserved quantities in terms of the standard ADM quantities as follows
\begin{align}
	\rho_E &= - \alpha T^0_0 = \alpha \rho - \beta_i S^i, \\
	\rho_{J} &= S_{\phi}, \\
    J_t^i &= \alpha\, \gamma^{ij}\left(\alpha S_j - \beta^k S_{jk}\right) - \beta^i \rho_{E}, \\
    J_{\phi}^i &= \alpha \gamma^{ij} S_{j\phi} - \beta^i \rho_J.
\end{align}
Another common quantity of interest is the ADM mass of the spacetime \cite{Intro3p1} defined as 
\begin{equation}
    M_{\textup{ADM}} = \frac{1}{16\pi}\lim_{r \to \infty}\int_{\partial \Sigma} (\partial_i h^{ij} - \partial^i h^j_j) \dd S_i,
    \label{eq:ADM_mass}
\end{equation}
where we assume that the boundary surface is in the weak field limit 
\begin{equation}
    g_{\mu\nu} = \eta_{\mu\nu} + h_{\mu\nu},
\end{equation}
where $h_{\mu\nu}$ is small. 
In our case if we include the backreaction we find
\begin{equation}
    h_{\mu\nu} = h^{BH}_{\mu\nu} + h^{\varphi}_{\mu\nu},
\end{equation}
where $h^{BH}_{\mu\nu}$ is from the BH background metric and $h^{\varphi}_{\mu\nu}$ from the scalar field backreaction. As \eqref{eq:ADM_mass} is linear in $h_{\mu\nu}$ these give seperable contributions to $M_{\textup{ADM}}$. Perturbing the Einstein equations to first order gives
\begin{equation}
\begin{split}
    8\pi T^i_t &=\tfrac{1}{2}(\partial_t [ \partial_j h^{ij} - \partial^i h^j_j] + \partial^2_t h^{it} + \partial^i \partial_{\mu} h^{\mu}_t - \partial_{\mu} \partial^{\mu} h^i_t) + \mathcal{O}(h^2), \\
    &= \tfrac{1}{2} \partial_t [ \partial_j h^{ij} - \partial^i h^j_j] + \mathcal{O}(h^2),
\end{split}              
\end{equation}
using the synchronous gauge $h_{\mu t}=0$. Given that the BH background is fixed in time, we then recover
\begin{equation}
    \partial_t M_{\textup{ADM}} = \lim_{r \to \infty}\int_{\partial \Sigma} \sqrt{-g} T^i_t \dd S_i + \mathcal{O}(h^2),
\end{equation}
where we can reintroduce $\sqrt{-g}$ as in the weak field limit $\sqrt{-g} = 1 + \mathcal{O}(h)$. Hence if we choose a large enough radius sphere the flux across the surface, and thus the change in the mass within the sphere \eqref{eq:change_sphere_mass}, approximates to the change in the ADM mass which would be measured. 

\begin{figure}
    \centering
    \includegraphics{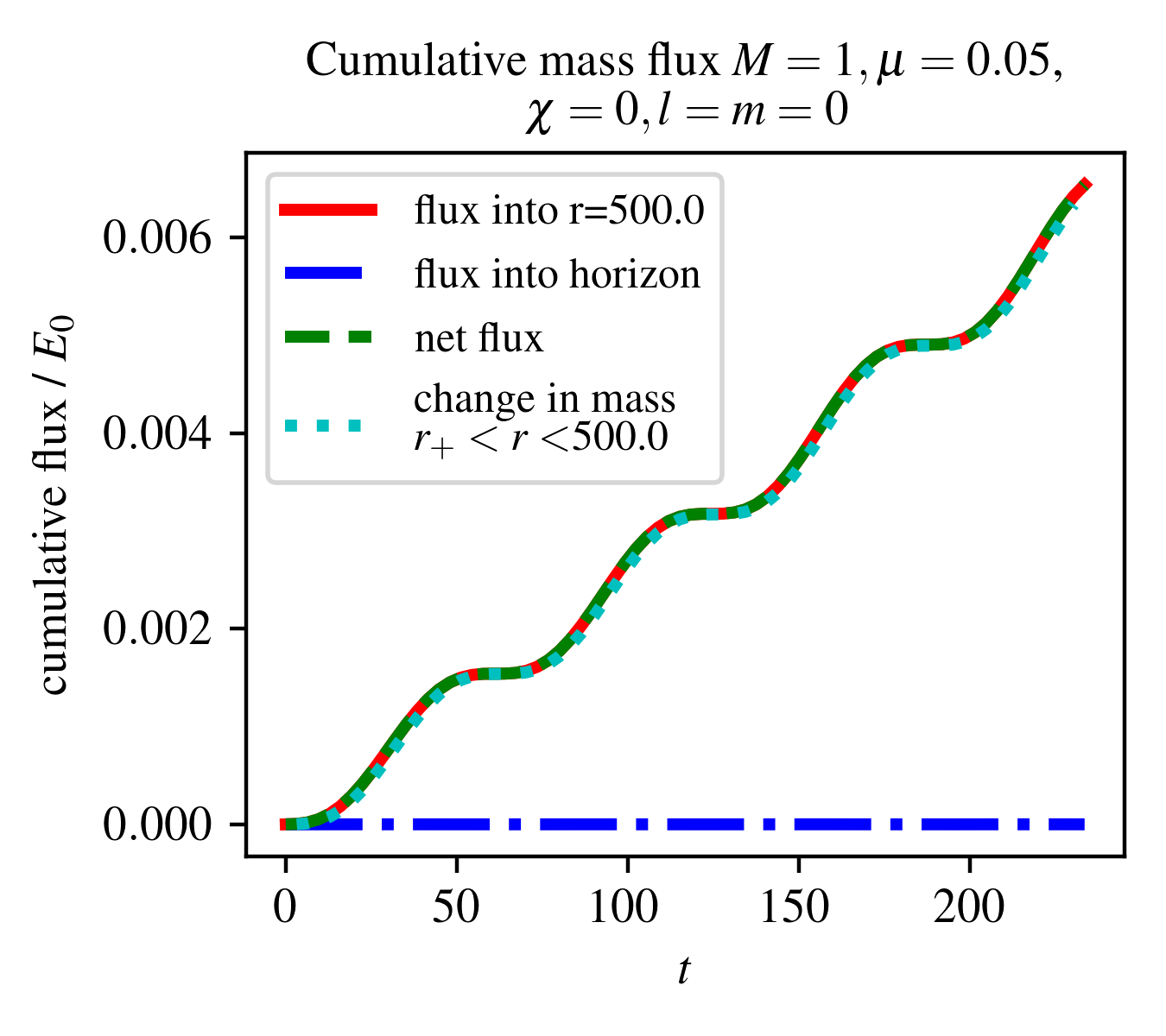}
    \caption{The cumulative mass flux into a sphere radius $R=500M$, into the horizon of the BH, the net flux and the change in cloud mass measured from integrating the density $\rho_E$, for $M\mu=0.05, \chi=l=m=0$ in Kerr-Schild coordinates. We see good agreement between the change in cloud mass and the net mass flux.}  \label{fig:KS_mass_flux_agreement}
\end{figure}

We have verified that the integral of the scalar field mass flux over a sphere does correspond to the change in scalar field mass inside the sphere in both quasi Isotropic Kerr coordinates (Fig. \ref{fig:BL_mass_flux_agreement}) and the alternative Kerr-Schild coordinates (Fig. \ref{fig:KS_mass_flux_agreement}). 

\subsection{Convergence tests}

\begin{figure}
    \centering
    \includegraphics[width=0.5\textwidth]{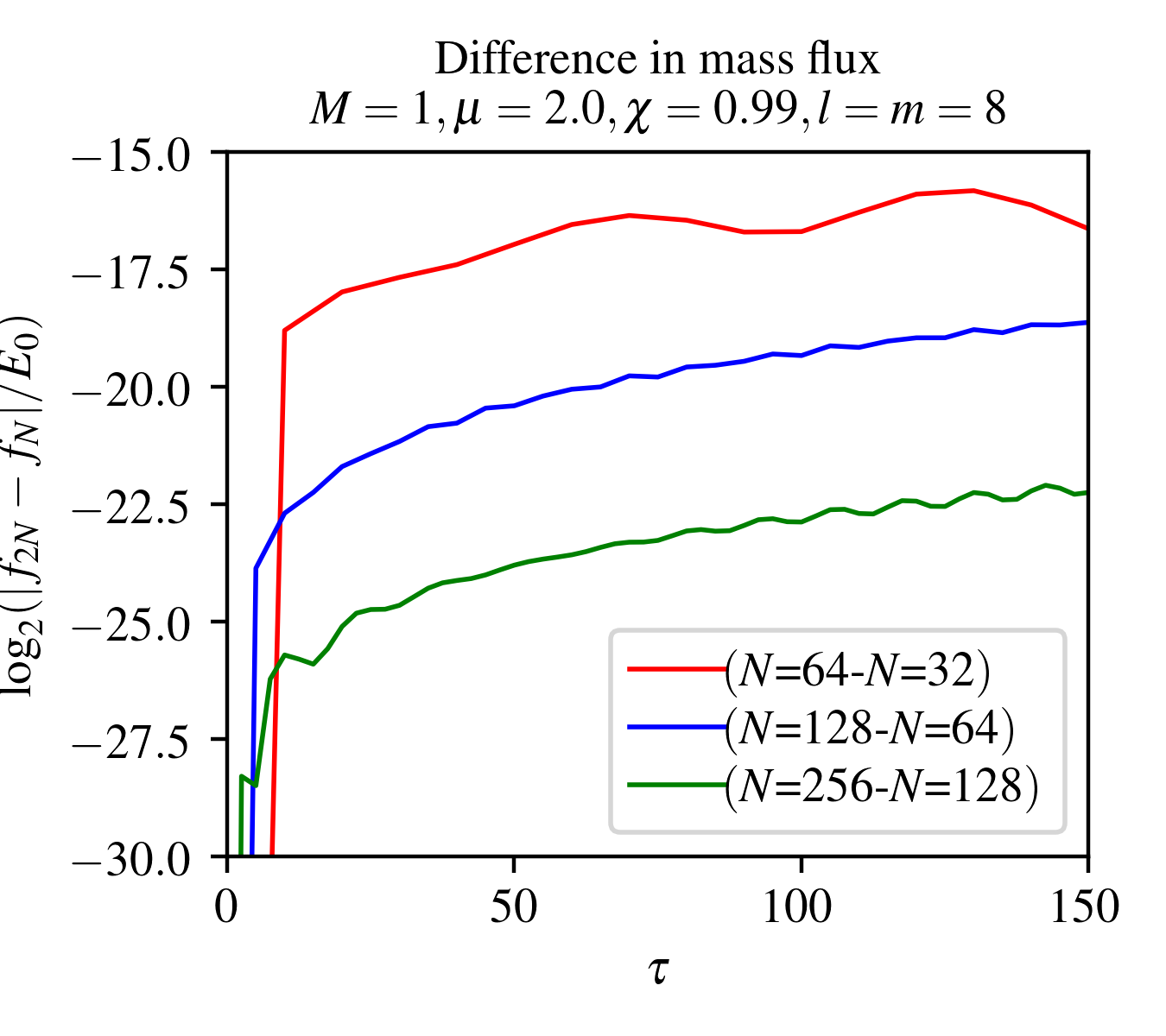}
    \caption{Difference in mass flux through $R=300M$ on doubling $N$ for $M\mu=2.0, \chi=0.99, l=m=8$ plotted on a log scale. We see the difference in flux decreases by a factor of $\sim 2^{-4}$ each time, indicating we do indeed have 4th order convergence.}  \label{fig:compare_N_convergence_diff}
\end{figure}

\begin{figure}
    \centering
    \includegraphics[width=0.5\textwidth]{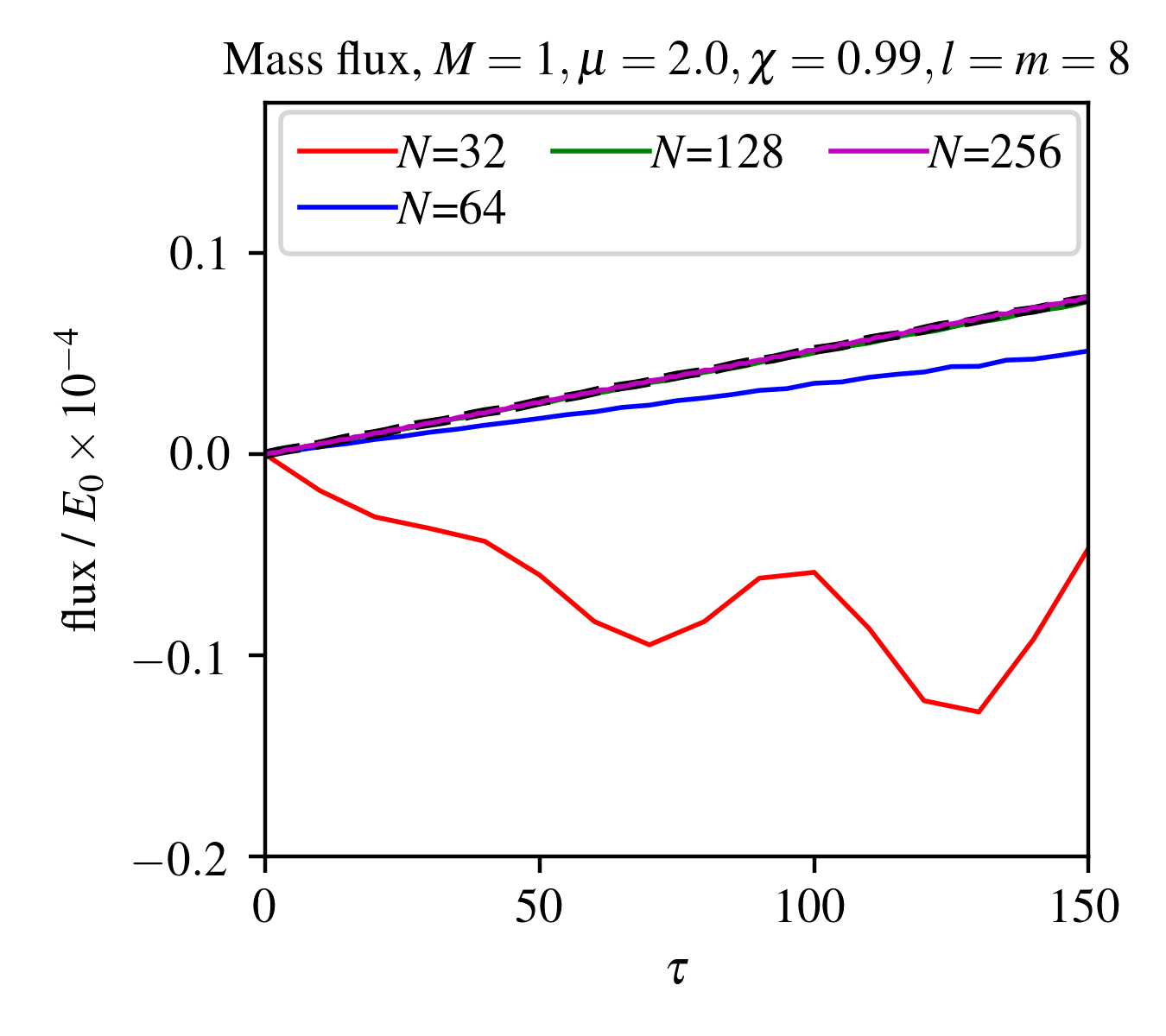}
    \caption{Mass flux through $R=300M$ for different $N$ and in the most challenging case studied $M\mu=2.0, \chi=0.99, l=m=8$. We see good agreement by eye for $N \geq 128$ which corresponds to the values used. The black dashed line is the perturbative analytic solution which we expect to diverge from the true numerical solution at later times, but provides a guide to the expected result at early times.}  \label{fig:compare_N_convergence}
\end{figure}

In this section we illustrate our tests of the numerical convergence of our code. As we use fourth order finite difference stencils to evolve the field, we expect our errors to decrease with $N$, the number of grid cells, as $N^{-4}$. The first quantity we test is the scalar field mass flux into a sphere of $R=300M$, a quantity we explored in Sec. \ref{sec-numerics}. We compute the flux for the most challenging case studied of high scalar mass $M\mu=2.0$, extremal spin $\chi=0.99$ and large scalar angular momentum $l=m=8$. The results for $N=32, 64, 128, 256$ are shown in Fig. \ref{fig:compare_N_convergence}. By eye we see good agreement for $N \geq 128$ which corresponds to the resolution used ($N=128$). To test the convergence we also plot the difference between the flux $f$ on doubling the resolution from $N$ to $2N$ (Fig. \ref{fig:compare_N_convergence_diff}). We would expect this difference to decrease by $\sim 2^{-4}$ on doubling $N$, and indeed this is approximately what we observe. 

\begin{figure}
    \centering
    \includegraphics{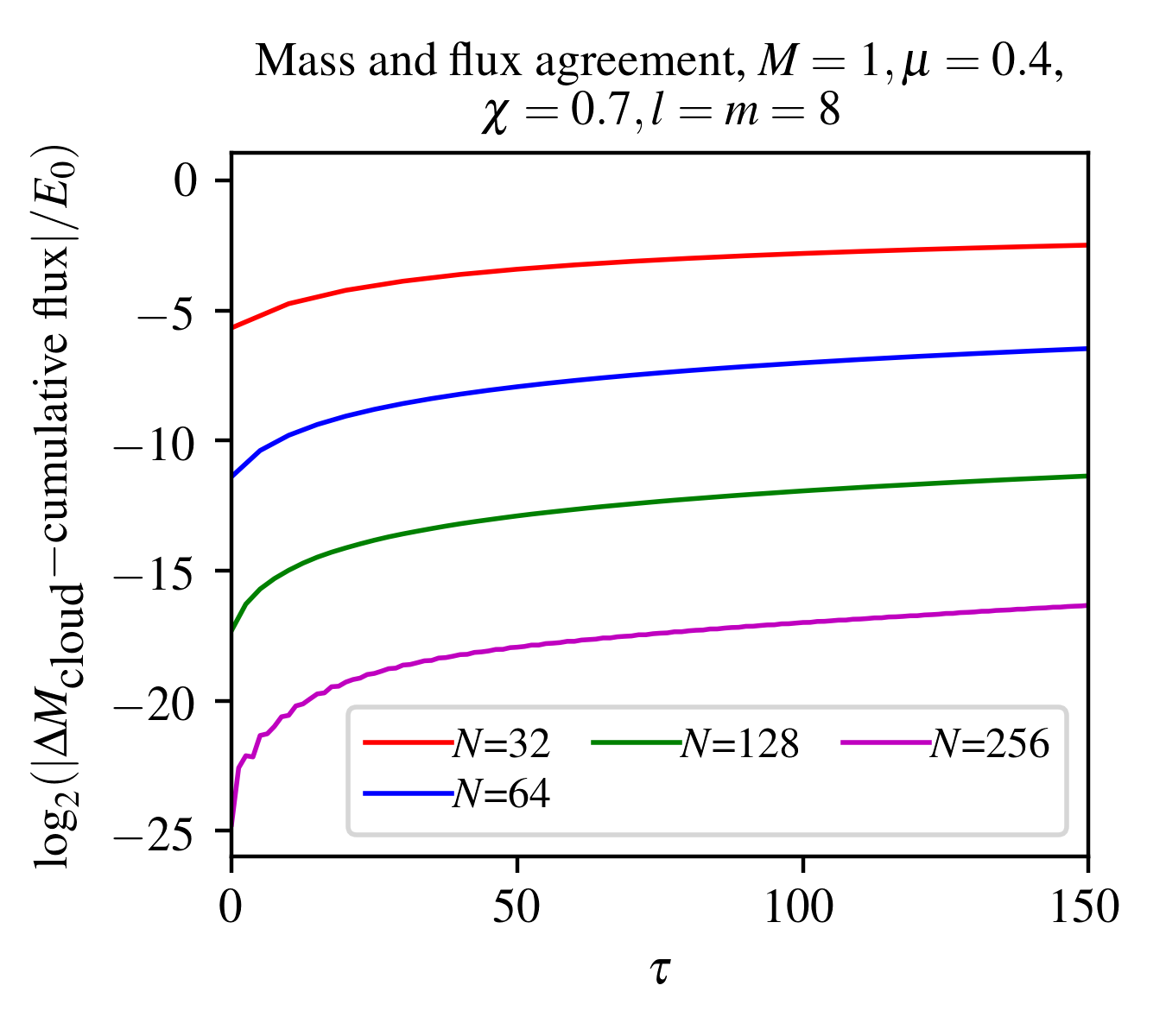}
    \caption{Difference between the change in mass inside a sphere radius $R=300M$ measured from integrating the density $\rho_E$, and the cumulative mass flux into the sphere. This difference should be zero for an infinite resolution simulation. We show values for $M\mu=2.0, \chi=0.99, l=m=8$ and different $N$ again plotted on a log scale. We see that again doubling $N$ decreases the error by a factor of $\sim 2^{-4}$ each time, indicating 4th order convergence. }
    \label{fig:BL_mass_flux_agreement}
\end{figure}

Finally, we examine the difference between the change in total mass inside a sphere radius $R=300$ and the cumulative flux into the sphere. As we established in Sec. \ref{sec:diagnostics} this difference should be zero. Fig. \ref{fig:BL_mass_flux_agreement} shows the difference in QIK coordinates for increasing N. We can see that the difference is both small for our typical choice of $N=128$ and it decreases by approximately $2^{-4}$ on doubling $N$ as we expected.

\FloatBarrier

\bibliography{biblio}

\end{document}